\documentclass[%
 aip,
amsmath,amssymb,
preprint,
]{revtex4-1}

\pdfoutput=1

\usepackage{graphicx}% Include figure files
\usepackage{dcolumn}% Align table columns on decimal point
\usepackage{bm}% bold math
\usepackage[utf8]{inputenc}
\usepackage[T1]{fontenc}
\usepackage{mathptmx}
\usepackage{etoolbox}

\usepackage[english]{babel}
\usepackage{textcomp} 

\usepackage{subcaption}

\usepackage{xcolor}
\usepackage{hyperref}
\hypersetup{
    colorlinks=true,
    linkcolor=blue,
    filecolor=magenta,      
    urlcolor=cyan,
}
\usepackage[capitalise]{cleveref}

\makeatletter
\def\@email#1#2{%
 \endgroup
 \patchcmd{\titleblock@produce}
  {\frontmatter@RRAPformat}
  {\frontmatter@RRAPformat{\produce@RRAP{*#1\href{mailto:#2}{#2}}}\frontmatter@RRAPformat}
  {}{}
}%
\makeatother
\begin{document}

\title[]{Hamiltonian estimation of island width threshold for stochasticity onset on edge pedestal top in presence of a resonant magnetic perturbation}
% Force line breaks with \\
\author{Zhifei Gui}
\affiliation{School of Physics, Huazhong University of Science and Technology, Wuhan 430074, China.}

\author{Ping Zhu}
\affiliation{State Key Laboratory of Advanced Electromagnetic Technology,
International Joint Research Laboratory of Magnetic Confinement Fusion
and Plasma Physics, School of Electrical and Electronic Engineering,
Huazhong University of Science and Technology, Wuhan 430074, China.}
\affiliation{Department of Nuclear Engineering and Engineering Physics, University of Wisconsin--Madison, Madison, WI 53706, USA.}
\email{zhup@hust.edu.cn}

\author{Dominique Franck Escande}
\affiliation{Aix-Marseille Universit\'e, CNRS, PIIM, UMR 7345, Marseille, France.}

\date{\today}
             
\begin{abstract}
This study applies the Hamiltonian method to analyze the nonlinear magnetic topology induced by Resonant Magnetic Perturbations (RMPs) in tokamaks. We investigate the system's chaotic behavior by comparing three methods: the renormalization method, Lyapunov exponents (LE), and weighted Birkhoff average (WBA). A strong consistency is found among these methods in predicting the large scale stochasticity threshold. The magnetic island width threshold provides a quantitative criterion for optimizing RMP-based ELM control, bridging a critical gap in plasma control strategies.
\end{abstract}

\keywords{tokamak, magnetic islands, resonant magnetic perturbations, Hamiltonian chaos, Hamiltonian mapping}

\maketitle
% ------------------------------------------------------------------------ %
\section{Introduction}
Resonant Magnetic Perturbations (RMPs) are a crucial tool for controlling edge localized modes (ELMs) in tokamaks~\cite{Evans2006}. However, the underlying nonlinear physics poses significant challenges for predictive modeling. Traditional linear MHD theories cannot accurately predict magnetic island widths or stochasticity onset, as those observed in DIII-D~\cite{Xiao2016,Xiao2017} and ASDEX Upgrade experiments~\cite{Willensdorfer2024}. The Hamiltonian dynamics~\cite{CARY1983,Pina1988} offers a powerful alternative. Subsequent developments, such as the renormalization group method for predicting chaos thresholds~\cite{Escande1981, Escande1981-ok, Escande1985, Chandre2002, Koch2004, Escande2018}, have provided robust tools. However, a comprehensive model for toroidal configurations that directly connects physical parameters to the chaos threshold has remained a key objective.

Over the past decades, a variety of chaos detection techniques have been developed and applied in Hamiltonian systems. Early methods, such as Lyapunov exponents~(LE)~\cite{BRIDGES2001219}, quantify sensitivity to initial conditions. More recent indicators include the Smaller Alignment Indices~(SALI)~\cite{Skokos2016}, which tracks the evolution of deviation vectors, and the Lagrangian Descriptor~(LD)~\cite{DAQUIN2022133520, Daquin2023}, which highlights phase space structures. While these methods are effective, they can be limited by computational cost, noise sensitivity, or difficulty distinguishing weak chaos. In this work, we use the weighted Birkhoff average~(WBA)~\cite{Sander2020-wo, Duignan2023,Ruth2024-rt}, which efficiently distinguishes regular and chaotic trajectories by analyzing the convergence of time-averaged observables. WBA offers super-exponential convergence for regular orbits and clear separation from chaotic ones, making it well suited for large-scale studies of magnetic field line stochasticity.

In this work, we address this gap by deriving a generalized Hamiltonian for toroidal magnetic field lines and subsequently transforming it into a standard two-wave form for renormalization analysis. This approach yields a chaos threshold criterion connecting measurable island widths to universal parameters. Here, the threshold for large scale stochasticity refers to the disappearance of the last KAM torus separating two main stochastic layers, beyond which magnetic field lines can access a wide region of phase space, although some magnetic islands may still persist. We use high-efficiency symplectic mapping techniques~\cite{Abdullaev2006ph} to simulate the system's dynamics. We also validate our threshold predictions by using LE and WBA. This comprehensive methodology yields a more precise and physically substantiated predictive capability compared to the conventional Chirikov criterion, thereby serving as a quantitative method for the optimization of RMP-based ELM control strategies.

The rest of the paper is structured as follows: 
\cref{sec:theory_and_model} presents the Hamiltonian method for magnetic field line motion and the symplectic mapping model. 
\cref{sec:chaotic_and_stochastic_criteria} introduces the renormalization method and other chaos diagnostic tools. 
\cref{sec:numerical_simulation_and_analysis} presents the numerical results, comparing the predictions of the different methods. 
\cref{sec:stochasticity_threshold_DIII-D_1} analyzes the stochasticity threshold of resonant and non-resonant magnetic islands induced by RMP in DIII-D. \cref{sec:stochasticity_threshold_DIII-D_nonresonant} further investigates the stochasticity threshold between non-resonant magnetic islands in DIII-D. 
Finally, \cref{sec:summary} summarizes the findings and discusses their implications.
% ------------------------------------------------------------------------ %
\section{Hamiltonian system of tokamak magnetic field lines}\label{sec:theory_and_model}
The trajectory of magnetic field lines in a tokamak can be described as a Hamiltonian system, with the toroidal angle $\varphi$ serving as the time-like variable. This method is a powerful tool for analyzing the topology of perturbed magnetic fields.
\subsection{Hamiltonian field line equations and toroidal Hamiltonian}
The Hamiltonian of magnetic field lines can be decomposed into an integrable part, denoted as $H_0$, and a non-integrable perturbation $H_1$. This is done by using action-angle variables $(\psi, \theta)$, where $\psi$ is the toroidal flux and $\theta$ is the poloidal angle.
The total Hamiltonian is expressed as
\begin{equation}\label{total_Hamiltonian_flux}
H(\psi,\theta, \varphi) = H_0(\psi) + \epsilon H_{1}(\psi,\theta, \varphi),  
\end{equation}
where $\epsilon$ is a small perturbation magnitude. And the perturbation term $H_1$ can be generally expressed as a Fourier series
\begin{equation}\label{eq:perturbation_expansion}
    H_1 = \sum_{m,n} H_{mn}(\psi)\cos(m\theta - n\varphi + \chi_{mn}),
\end{equation}
where $(m,n)$ are the poloidal and toroidal mode numbers, respectively. The canonical equations are
\begin{equation}\label{eq:canonical_eq_simple}  
\frac{\mathrm{d}\theta}{\mathrm{d} \varphi}=\frac{\partial H}{\partial \psi}, 
\quad \frac{\mathrm{d} \psi}{\mathrm{d} \varphi}=-\frac{\partial H}{\partial\theta}.  
\end{equation}

For analytical developments it is convenient to use action-angle variables $(J,\theta)$ with $J=r^2/2$ and $\psi = J/R_{0}^2$ in the large aspect ratio approximation~\cite{RLVianaHamiltonian2023,Abdullaev2014}, where the cylindrical-like coordinate system $(r,\theta,\varphi)$ is adopted, with $r$ being the radial distance to the magnetic axis and $R_{0}$ being the major radius of the tokamak. Then the field-line Hamiltonian (see \cref{APP.A} for details) reads
\begin{equation}
\label{eq:generalized_toroidal_Hamiltonian}
H(J,\theta,\varphi)
= H_0(J) + \epsilon \sum_{j=1}^{2} V_j(J)\cos\Phi_j,
\end{equation}
where $\Phi_j \equiv m_{j}\theta - n_j\varphi$, 
\begin{equation}
H_0(J)=\int^{J}\frac{\mathrm{d}J'}{q(J')}, 
\end{equation}
and
\begin{equation}
V_j(J)=\big|I_j(J)\big|,
\qquad 
I_j(J)=\int^{J}\frac{B_{1\theta}^{(j)}(\sqrt{2J'})}{\sqrt{2J'}}\,\mathrm{d}J'.
\end{equation}

\subsection{The paradigm Hamiltonian}
Following the calculation in \cref{APP.A}, the Hamiltonian~\eqref{eq:generalized_toroidal_Hamiltonian} is transformed into the two-wave paradigm form (section 2.1 of Ref.~[\onlinecite{Escande1985}])
\begin{equation}\label{eq:paradigm_H}
H(X,Y,Z)=\frac{1}{2}X^2-M\cos Y-P\cos[k(Y-Z)],
\end{equation}
where $X$, $Y$, and $Z$ are normalized canonical variables. The wave number ratio is $k = m_2 / m_1$. For a single $(m_{j},n_j)$ harmonic, the magnetic island width at the rational surface $r_s$~(see Eq.~(8.55) of Ref.~[\onlinecite{Zohm2014-yc}]) is
\begin{equation}\label{eq:magnetic_island_width}
W_j = 4 \sqrt{\frac{B_{1r}^{(j)}(r_s)\; r_s\; q(r_s)}{m_{j}\; q^{\prime}(r_s)\; B_{0\theta}(r_s)}}.
\end{equation}
The normalized strengths in the paradigm Hamiltonian $M$ and $P$ can be written in terms of island widths, i.e.
\begin{equation}\label{island_width_MP_relation}
\begin{aligned}
M &\simeq \epsilon\,\frac{q'(r_s)^{2}}{16\, q(r_s)^{4}}\; 
\frac{W_1^2}{\left(\frac{n_{2}}{m_{2}}-\frac{n_{1}}{m_{1}}\right)^2},\\
P &\simeq \epsilon\,\frac{q'(r_s)^{2}}{16\, q(r_s)^{4}}\; 
\frac{W_2^2}{\left(\frac{n_{2}}{m_{2}}-\frac{n_{1}}{m_{1}}\right)^2} .
\end{aligned}
\end{equation}
These expressions provide a direct conversion from experimentally accessible island widths to the parameter space $(M,P)$ used by the renormalization analysis.

\subsection{Hamilton-Jacobi mapping}\label{sec:hamilton_jacobi_mapping}
To numerically investigate the dynamics of the Hamiltonian system described by \cref{total_Hamiltonian_flux}, we employ an efficient symplectic mapping method developed by Abdullaev (see Chapter 4 of Ref.~[\onlinecite{Abdullaev2006ph}]).
This approach reduces computational cost and long-time error accumulation relative to small-step integrators. It is therefore ideal for studying long-term chaotic behavior, as it advances the system over large steps in the toroidal angle.

The method constructs a discrete map $F:(\psi, \theta) \mapsto (\overline{\psi},\overline{\theta})$ that preserves the symplectic nature of the Hamiltonian flow. The mapping equations are defined implicitly through a generating function $S$, which is constructed from the perturbation Hamiltonian $H_1$ (\cref{eq:perturbation_expansion}):
\begin{equation}
\label{Abdullaev_map}
\begin{cases}
\overline{\psi} & = \psi-\epsilon\frac{\partial S}{\partial\theta}\left(\theta,X,\varphi\right)-\epsilon\frac{\partial S}{\partial\overline{\theta}}\left(\overline{\theta},X,\overline{\varphi}\right)\\
\overline{\theta} & = \theta + \frac{2\pi}{N_{\mathrm{step}}}\Omega\left(X\right) + \epsilon\frac{\partial S}{\partial X}\left(\theta,X,\varphi\right) + \epsilon\frac{\partial S}{\partial X}\left(\overline{\theta},X,\overline{\varphi}\right) \\
\overline{\varphi} & = \varphi + \frac{2\pi}{N_{\mathrm{step}}}
\end{cases}
\end{equation}
with the step $\Delta \varphi = 2\pi/N_{\mathrm{step}}$. The auxiliary variables are defined as
$X=\psi-\epsilon \partial S\left(\theta,X,\varphi\right)/\partial\theta$
and
$\Omega\left(\psi\right)=1/q\left(\psi\right)$. And the generating function takes the form
\begin{equation}\label{generating function}
\begin{aligned}
S\left(\theta,X,\varphi\right)
&=\frac{\pi}{N_{\mathrm{step}}}\sum_{m,n}H_{mn}\left(\psi\right)\\
&\times \left[a\left(x_{mn}\right)\sin\left(m\theta-n\varphi\right)
+ b\left(x_{mn}\right)\cos\left(m\theta-n\varphi\right)\right]
\end{aligned}
\end{equation}
with coefficient functions are
\begin{equation}
a\left(x\right)=\frac{1-\cos x}{x},\quad b\left(x\right)=\frac{\sin x}{x},\quad x_{mn}=\frac{\pi}{k}\left(m\Omega\left(X\right)-n\right).
\end{equation}
Using Escande's magnetic island width formula from Section 5.3 of Ref.~[\onlinecite{Escande2024}]
\begin{equation}\label{eq:Escande_island_width}
W=4\sqrt{\left|\frac{H_{1,m_0,n_0}(p_0)}{\frac{\mathrm{d}t(p_0)}{\mathrm{d}p}}\right|},
\end{equation}
the magnetic island width $W$ can be computed by substituting $H_{mn}$ from \cref{eq:perturbation_expansion}. The $t$ is the rotational transform and $p_{0}$ is the toroidal flux $\psi_t$ defined by $t(p_0)=n_0/m_0$.

Using the trajectories generated by the mapping method described in \cref{sec:hamilton_jacobi_mapping}, we identify and quantify the system's chaotic behavior with the criteria introduced in \cref{sec:chaotic_and_stochastic_criteria}.
% ------------------------------------------------------------------------ %
\section{Chaotic and Stochastic Criteria for Magnetic Surfaces}\label{sec:chaotic_and_stochastic_criteria}
We employ three independent criteria to diagnose and quantify the system's chaotic behavior: 1) the {renormalization method}~\cite{Escande1985}, which predicts the breakup of KAM tori by iteratively analyzing the stability of the two main resonances in the paradigm Hamiltonian; 2) the LE method~\cite{BRIDGES2001219}, which measures the chaos magnitude by calculating the separation rate of adjacent orbits; and 3) the WBA method~\cite{Sander2020-wo, Duignan2023,Ruth2024-rt}, which distinguishes regular from chaotic orbits by analyzing the convergence of their time series.
\subsection{Renormalization Method}
The stability of KAM tori for the paradigm Hamiltonian, 
$H(v,x,t) = v^2/2 - M \cos x - P \cos k(x - t)$, 
can be analyzed using the renormalization method~(for more details see Ref.~[\onlinecite{Escande1985}]). 
This method recursively maps the original system to a new one with the same form 
but with renormalized parameters $(M^{\prime}, P^{\prime})$. 
The recurrence relations for the parameters are given by
\begin{equation}
\begin{aligned}
M^{\prime}&=B_{m}(k)M^{l}P[1+\mathcal{O}(M^{2})],\\
P^{\prime}&=C_{m}(k)M^{m}P[1+\mathcal{O}(M^{2})],
\end{aligned}
\end{equation}    
where $B_m(k)$ and $C_m(k)$ are coefficients dependent on the wave number ratio $k$. And $m=\operatorname{Int}(z),\quad l=m-1$ where $z$ is the zoning number of a torus $\mathcal{T}$ between the two main resonances. The stability of a KAM torus is determined by the behavior of the sequence $\{M^{(i)},P^{(i)}\}$ under this iterative mapping. If the sequence converges, the torus is stable; if it diverges, the torus has broken, indicating a transition to chaos.

Renormalization can be equivalently formulated in terms of the parameter triple $(M, P, k)$, or in terms of $(P, M, 1/k)$ after exchanging the two resonances. In the following content, we adopt the convention that $M$ denotes the amplitude of the resonance treated as the reference pendulum in a given renormalization step, and $P$ denotes the secondary wave; when the alternative ordering is used, we will state it explicitly.

To locate the most robust torus between two resonances $(m_1,n_1)$ and $(m_2,n_2)$, we use a noble-number construction. Let $u_1=n_1/m_1$ and $u_2=n_2/m_2$, and order them so that $u_{\min}<u_{\max}$. We build a target irrational rotation number $\rho\in(u_{\min},u_{\max})$ by taking the longest common prefix of the continued-fraction expansions of $u_{\min}$ and $u_{\max}$, and appending a tail of ones $[1,1,1,\dots]$. This noble $\rho$ is maximally Diophantine and thus corresponds to the most robust torus between the two considered resonances. The same $\rho$ is then used as the target rotation number when computing LE and WBA on that torus.

\subsection{Lyapunov Exponent Method}  
The LE method quantifies the average exponential rate of separation of infinitesimally close trajectories in phase space. For a discrete map $\mathbf{x}_{k+1} = \mathbf{F}(\mathbf{x}_k)$, where $\mathbf{x}_k = (\psi_k, \theta_k)^T$, the maximal LE, $\lambda_1$, characterizes the system's predictability. A regular, quasi-periodic orbit on a KAM torus has $\lambda_1 = 0$, whereas a chaotic orbit exhibits exponential divergence of trajectories, resulting in a positive maximal LE, $\lambda_1 > 0$. We use a standard algorithm based on QR decomposition (where "QR" refers to the factorization of a matrix into an orthogonal matrix $Q$ and an upper triangular matrix $R$) to compute the LE for trajectories generated by the symplectic map~\cite{BRIDGES2001219}.

\subsection{Weighted Birkhoff Average Method}  
The WBA method provides a robust method to distinguish between quasi-periodic and chaotic orbits based on their convergence properties. For an observable $h(\psi, \theta)$, the WBA over $T$ iterations is defined as
\begin{equation}
\mathcal{WB}_T(h)(\psi_0,\theta_0)
=\sum_{k=0}^{T-1}w\left(\frac{k}{T}\right)h(\psi_k,\theta_k)/
\sum_{k=0}^{T-1}w\left(\frac{k}{T}\right),
\end{equation}
where $w(t)$ is a smooth weight function, $k$ is the iteration index, and $h(\psi_k,\theta_k)$ is the value of observable at $k$th step. For regular orbits, the WBA converges super-exponentially fast to the true average, while for chaotic orbits, the convergence is much slower.

In our numerical implementation, we use the $C^{\infty}$ compactly supported weight function $w(t)=\exp\left(-[t(1-t)]^{-1}\right)$ and $h(\psi,\theta)=\psi$. To quantify the convergence rate and classify orbits, we use a digit accuracy, $\mathrm{dig}_T$, based on the residual between the WBAs calculated over the first and second halves of an orbit of total length $2T$~\cite{Sander2020-wo}
\begin{equation}\label{eq:digit accuracy}
\mathrm{dig}_T=-\log_{10}|\mathcal{WB}_T(h)(\psi_0,\theta_0)-\mathcal{WB}_T(h)(\psi_T,\theta_T)|.
\end{equation}
Regular orbits have a smooth dynamical structure, and their rotation numbers are Diophantine numbers. $\mathcal{WB}_T$ exhibits superconvergence for such orbits, with errors decreasing faster than any power of the number of iterations $T$ as $T$ increases. However, chaotic orbits lack smoothness, and the convergence rate of $\mathcal{WB}_T$ is very slow. Therefore, a high $\mathrm{dig}_T$ value indicates a fast convergence rate and a regular orbit, whereas a low $\mathrm{dig}_T$ value signifies a slow convergence rate and a chaotic orbit. Based on numerical experiments, we set a threshold $\mathrm{dig}_T < 5$ as the criterion for identifying a chaotic orbit. This distinction allows for a clear and efficient classification of orbits in phase space.
% ------------------------------------------------------------------------ %
\section{Onsets of magnetic chaos in response to RMP}\label{sec:numerical_simulation_and_analysis}
We consider two types of perturbations as in Ref.~[\onlinecite{Abdullaev2014},\onlinecite{RMPandmapping}]: MHD-type~(internal instability or response)~and RMP-type~(external coil field)~, as illustrated in \cref{fig:RMP perturbation and MHD perturbation diagrams}. The MHD perturbation dominates near the resonance, which can be modeled as
\begin{equation}\label{MHD perturbation}
H_{mn}(\psi)=\frac{n}{m} \left[\frac{\left(\frac{\psi}{\psi_{mn}}\right)^{1/\Delta }+\left(\frac{\psi}{\psi_{mn}}\right)^{-1/\Delta}}{2}\right]^{\frac{-m}{2}\Delta},
\end{equation}
whereas the RMP perturbation mainly excites the edge chaotic layer due to its exponential decay, which takes the form
\begin{equation}\label{RMP perturbation}
H_{mn}(\psi)=\frac{n}{m}\psi^{m/2}e^{-mC_0\psi W(\psi)/\gamma}.
\end{equation}

Simulation parameters are set as follows: $C_0 = 7\times10^{-2}$, $\gamma = 0.21$, $q(\psi)=1.25(1 + \psi)$, $\psi=\psi_t/\psi_a$, minor radius $a = 0.5~\mathrm{m}$, major radius $R_0 = 5~\mathrm{m}$. Initial points are 100 points, with $N=5 \times 10^3$ iterations. These parameters are chosen to be representative of typical large aspect ratio tokamak edge conditions and are consistent with previous studies~\cite{Abdullaev2014, RMPandmapping}. The values of $C_0$ and $\gamma$ characterize the spatial decay and scaling of the RMP field, while the linear $q$-profile is adopted for analytical simplicity. The resulting perturbation profiles $H_{mn}(\psi)$ for both RMP and MHD cases, calculated using these parameters, are plotted in~\cref{fig:RMP perturbation and MHD perturbation diagrams}. The safety factor profile $q(\psi) = 1.25(1 + \psi)$ is shown in~\cref{fig:q-profile}. For the (2,1) RMP perturbation, since~\cref{RMP perturbation} only describes the vacuum response, we superimpose $1/10$ of the MHD perturbation onto the RMP perturbation to mimic the resonant plasma response at $q=2/1$. The factor $1/10$ is chosen as a representative value to qualitatively account for the reduced amplitude of the plasma response compared to the vacuum RMP field near the resonance. For the (3,2) mode, only the MHD perturbation is applied at $q=3/2$. Owing to the linearity of the generating function~\eqref{generating function}, the magnitudes $\epsilon_{2,1}$ and $\epsilon_{3,2}$ in mapping~\eqref{Abdullaev_map} can be independently adjusted. For simplicity, we set $\epsilon_{2,1}=\epsilon_{3,2}$.

\subsection{Results from renormalization method}
We use the renormalization method to analyze the stability of quasiperiodic tori in the $(M, P)$ parameter space. Numerical results show clear convergence and divergence regions, corresponding to stable and unstable tori, respectively. After the transformation $x = 2\sqrt{M},\, y = 2\sqrt{P}$, \cref{fig:renormalization_combined} displays the convergence and divergence regions. To quantitatively determine the divergence boundary in the renormalization map, we employ a Support Vector Machine (SVM) classifier. The SVM is trained on numerically identified convergent and divergent points in the $(2\sqrt{M}, 2\sqrt{P})$ parameter space, and learns the optimal nonlinear boundary separating stable (convergent) and unstable (divergent) regions. The resulting SVM boundary agrees with the empirical criterion $MP^{\mathrm{g}-1} = 0.003$ from Ref.~[\onlinecite{Escande1985}] when one sets $k=1$ in \cref{eq:paradigm_H}, where $\mathrm{g} = (\sqrt{5} + 1)/2$ is the golden ratio. As $\epsilon$ increases, the system crosses from the convergence to the divergence region, indicating the onset of the large scale stochasticity. By means of the bisection method, we determined the critical perturbation magnitude $\epsilon_c \approx 0.0142$ in the renormalization method. 

In the following, we select three sample points for demonstration, whose respective perturbation magnitudes are $\epsilon_{2,1}=\epsilon_{3,2}=0.010$, $0.014$, and $0.020$. Based on the renormalization transformation defined in\cref{island_width_MP_relation}, the magnetic island widths can be transformed into the parameter space $(2\sqrt{M},2\sqrt{P})$. The pairs of points under different perturbation magnitudes in the parameter space are $(0.235,0.370)$, $(0.281,0.443)$, and $(0.338,0.531)$ respectively, and these sample points are shown in \cref{fig:renormalization_combined} with red circles. This is confirmed in the Poincaré plots (\cref{fig:poincare_maps}) obtained directly using the Abdullaev mapping from \cref{Abdullaev_map}. \cref{fig:renormalization_combined} shows that when $\epsilon=0.010~\text{or}~0.014$, the renormalized parameters are within the convergence domain, indicating convergence to the origin during iteration. However, when $\epsilon$ increases to $0.020$, the sample point crosses the critical boundary into the divergence domain. Then the iteration process exhibits  divergence in numerical simulation, signifying the system's transition to a state of the large scale stochasticity. As the perturbation magnitude increases, the system evolves from nested tori to large scale stochasticity, with magnetic islands expanding and overlapping (\cref{fig:poincare_maps}).

\subsection{Results from Lyapunov exponent method}
For scenarios with varying perturbation magnitudes, we systematically analyzed the evolution of the LEs of the Abdullaev map $F:\left(\psi, \theta\right) \mapsto \left(\overline{\psi}, \overline{\theta}\right)$ in relation to the number of iterations of magnetic field lines and their distribution characteristics in the initial magnetic flux $\psi$. As shown in \cref{fig:le_evolution}, both the maximal and average Lyapunov exponents increase significantly with the perturbation magnitude. The proportion of chaotic trajectories (i.e. growing gray curves) also increases with the perturbation magnitude $\epsilon$, and their LEs attain progressively larger values. This indicates that the system transitions from predominantly regular to increasingly chaotic behavior as the perturbation amplitude increases.

\subsection{Results from weighted Birkhoff average method}
The WBA method effectively distinguishes regular from chaotic orbits, allowing for a clear visualization of the phase space topology. \cref{fig:wba_analysis_combined} presents this analysis for increasing perturbation magnitudes. The left column (a)(c)(e) shows Poincaré plots where chaotic orbits (blue, identified by using digit accuracy $\mathrm{dig}_T < 5$) progressively expand and merge, destroying the regular KAM tori (black) and leading to large scale stochasticity at $\epsilon=0.020$. The right column (b)(d)(f) validates this classification by showing the convergence of WBA residuals. Regular orbits exhibit super-exponential convergence, while chaotic orbits converge slowly, consistent with the $\mathcal{O}(N^{-1/2})$ scaling. This clear separation in convergence rates confirms that WBA is a robust diagnostic of the system's nonlinear dynamics.

\subsection{Comparisons among renormalization, WBA and LE results}
To demonstrate the consistency between LE and WBA diagnostics, we plot the final LE versus the convergence digits $\mathrm{dig}_T$ for each trajectory under different perturbation magnitudes, as shown in \cref{fig:digT_vs_LE}.
The results show two clearly separated clusters: regular orbits have high $\mathrm{dig}_T$ and nearly zero LE, whereas chaotic orbits have low $\mathrm{dig}_T$ and positive LE. As $\epsilon$ increases, more orbits transition from regular to chaotic, confirming the equivalence and reliability of both diagnostics for identifying the stochasticity threshold.

The critical perturbation threshold $\epsilon_c$ for large scale stochasticity is determined by the destruction of the last KAM torus (golden mean torus), as identified using both LE and WBA methods. And the results are shown in \cref{fig:golden_mean_thresholds}. The left panel shows the maximal LE of the golden mean torus as a function of $\epsilon$. The LE exhibits a sharp increase at the threshold. Using the criterion $\lambda > 10^{-2}$ for chaos, we identify the critical magnitude at $\epsilon_c \approx 0.0139$. The right panel shows the WBA digit accuracy, $\mathrm{dig}_T$. A sudden drop in $\mathrm{dig}_T$ from a high value (regular orbit) to a low one (chaotic orbit) signals the torus breakup. This transition occurs between $\epsilon = 0.01388$ and $\epsilon = 0.01402$. Both results from WBA and LE are close to the result from renormalization that is $\epsilon_c \approx 0.0142$.
The strong agreement across three methods provides a robust and reliable determination of the threshold for the onset of large scale stochasticity in the system. 
% ------------------------------------------------------------------------ %
\section{Stochasticity threshold between resonant and non-resonant magnetic islands in a DIII-D experiment}\label{sec:stochasticity_threshold_DIII-D_1}
We consider a $q$-profile from the DIII-D experiment for ELM suppression which is a function of poloidal magnetic flux $\psi_p$ as~\cite{Abdullaev2014}
\begin{equation}\label{DIII-D q-profile}
q(\psi_p)=\frac{1}{2\pi\gamma}\ln\frac{Q}{|1-\psi_p|}+a_1(1-\psi_p)+a_2(1-\psi_p)^2,
\end{equation}
with parameters $\gamma=0.21$, $Q=9.8$, $q_0=1.1$, $q_{95}=3.92$ shown in \cref{fig:q_factor_DIII-D}. The parameters may be selected by calibrating them to align with the safety factor from experimental data or equilibrium code computations, using the formula \cref{DIII-D q-profile}. We let $a_1=0$ and $a_2=0$~\cite{Abdullaev2014}. From \cref{DIII-D q-profile}, the relation between the poloidal flux $\psi_p$ and the toroidal flux $\psi$ is given by
\begin{equation}
\begin{aligned}
\psi(\psi_p)&=\int_0^{\psi_p} q({\psi_p})\mathrm{d}{\psi_p}\\
&=\psi_a+\frac{1-{\psi_p}}{2\pi\gamma}\left(\ln\frac{Q}{|1-{\psi_p}|}+1\right)\\
&+\frac{a_1}{2}(1-{\psi_p})^2+\frac{a_2}{3}(1-{\psi_p})^3.
\end{aligned}
\end{equation}
where $\psi_a=2$ is the toroidal magnetic flux at the separatrix.

We assume a RMP with helicity $(3,1)$. The perturbation associated with MHD modes which influences a rational surface within the stochastic region defined by $\psi \in (0.8, 1)$. Both the RMP and MHD perturbations are assumed to have equal perturbation magnitudes, denoted as $\epsilon$. Using the same perturbation models Eqs.~(\ref{MHD perturbation}) and~(\ref{RMP perturbation}), the Hamiltonian of the system is 
\begin{equation}
\begin{aligned}
H &= \int \frac{1}{q(\psi)} \mathrm{d}\psi \\
&+ \epsilon \left[ H_{RMP}(\psi)\cos(3\theta-\varphi)+H_{MHD}(\psi)\cos(m\theta-n\varphi)\right].
\end{aligned}
\end{equation}
These parameters can be converted to the renormalization coefficients $(M, P)$ with the transformation relationship
\begin{equation}\begin{aligned}
M = \epsilon \frac{{\mathrm{d}q^{-1}}/{\mathrm{d}\psi}|_{31}}{(\Delta v)^2}H_{RMP}(\psi_{31}),\\
P = \epsilon \frac{{\mathrm{d}q^{-1}}/{\mathrm{d}\psi}|_{31}}{(\Delta v)^2}H_{MHD}(\psi_{mn}),
\end{aligned}\end{equation}
where $\Delta v = v_1 - v_2 = (m_1/q_1 - n_1) - (m_2/q_2 - n_2)$. We use the renormalization method to determine the critical perturbation threshold $\epsilon_c$ for each rational surface. For simplicity, a circular cross-section is assumed, with the coordinate transformation~\cite{Abdullaev2014}
\begin{equation}
\psi = 1-\left(1-r^2/R_0^2\right)^{1/2}
\end{equation}
where $R_0 = 1.69~\mathrm{m}$. 

\cref{fig:RMPvsMHD} shows that the critical width of RMP islands ($1.03$--$4.04\mathrm{cm}$, averaged width $2.46\mathrm{cm}$) is generally larger than that of MHD islands ($0.54$--$1.04\mathrm{cm}$, averaged width $0.82~\mathrm{cm}$). RMP-induced islands provide a background that facilitates the onset of stochasticity by activating smaller MHD islands at the plasma edge.
We choose the (4,1) and (7,2) perturbation case in MHD and use the WBA method to verify the chaos threshold obtained from the renormalization method.
The $\mathrm{dig}_T$ evolution with normalized $\epsilon$ is shown in the \cref{fig:digT_vs_epsilon_all}.
Initially, under conditions of minimal perturbation, we identified the most robust flux surface by analyzing the values of the convergence digit $\mathrm{dig}_T$. Subsequently, we computed $\mathrm{dig}_T$ within the normalized interval [0.5, 1.5] of the critical value $\epsilon_c$ in each case. 
It is evident that when the perturbation $\epsilon$ is significantly lower than the critical value $\epsilon_c$, $\mathrm{dig}_T$ remains at a relatively high level, suggesting the persistence of the most robust flux surface during this period. However, as $\epsilon$ approaches $\epsilon_c$, $\mathrm{dig}_T$ drops sharply, crossing our chaos criterion of $\mathrm{dig}_T=5$. This indicates the breakup of the most robust flux surface. % critical perturbation regime.
This declining trend demonstrates the breakup of the most robust flux surface in response to the critical perturbation. The critical magnetic islands at each non-RMP-resonant surfaces near edge are denoted on the q-profile~(\cref{fig:q_profile_islands}).

The Chirikov overlapping parameter $S$ is defined as
\begin{equation}
S=\frac{1}{2}\frac{W_{mn}+W_{31}}{\psi_{mn}-\psi_{31}}
\end{equation}
where $W_{mn}$ and $W_{31}$ represent the MHD island width and RMP island width, respectively. 
$\psi_{mn}-\psi_{31}$ denotes the distance between the centers of two magnetic surfaces. Typically, the Chirikov criterion predicts that a value of S that is greater than 1 indicates that neighboring magnetic islands are overlapping, which could lead to the stochasticity of magnetic field lines. However, our results show that the stochastic threshold occurs at much lower Chirikov parameter values than the $S=1$ criterion (\cref{fig:Chirikov}). The predicted critical island averaged widths ($1.7~\mathrm{cm}$ for RMP island, $0.58~\mathrm{cm}$ for MHD island) are consistent with recent experimental observations on ASDEX Upgrade~\cite{Willensdorfer2024}, supporting the validity of our approach.

In \cref{fig:Chirikov}, the minimal threshold for the onset of chaos lies between the $(3,1)$ RMP island and the $(4,1)$ MHD island. In this case, the RMP island corresponds to $M$ and the MHD island corresponds to $P$ in the renormalization method, with the renormalization procedure starting from $(M, P, k)$ with the wave number ratio $k$. When $P = 0$, the system reduces to a simple pendulum (an integrable system), so no chaos occurs and $S \to \infty$. When $P/M \sim 1$, the system is a non-integrable system. However, if $k \to 0$, the MHD island can be regarded as a fast perturbation, which can be eliminated by averaging, and at this time $S$ again approaches infinity. Therefore, the threshold is minimal near $k = M/P = 1$, because the coupling between the two resonances is the strongest at this point (see Fig.2.19 in Ref.~[\onlinecite{Escande1985}]).
% ------------------------------------------------------------------------ %
\section{Stochasticity threshold between non-resonant magnetic islands in DIII-D}\label{sec:stochasticity_threshold_DIII-D_nonresonant}

The two edge MHD islands may interact significantly and reach the stochasticity threshold before each of them interacts with the central $(3,1)$ RMP island. In this section, we further investigate the stochasticity threshold that results from the interaction between two non-resonant MHD magnetic islands at the plasma edge in DIII-D. In our analysis, the RMP-induced $(3,1)$ magnetic island is fixed. It is located radially closer to the tokamak core than the other islands. We have neglected the direct influence of the central $(3,1)$ RMP island and focus on the mutual interaction between two edge MHD islands, labeled as $(m_1, n_1)$ and $(m_2, n_2)$.

We apply the same renormalization procedure as in the previous section. \cref{fig:nonresonant_MHD_scatter} shows that the stochasticity threshold arising from the mutual interaction of two non-resonant edge MHD islands (mean critical width $\approx 0.33\mathrm{cm}$) is substantially lower than the threshold of interaction between an edge MHD island and the central RMP island. This indicates that couplings among edge MHD islands alone can produce large-scale stochastic layers form before the central RMP island becomes dominant.
% --------------------------------Summary-------------------------------- %
\section{Summary}\label{sec:summary}
In this work, we present a Hamiltonian method to a quantitative prediction of the stochasticity threshold of RMP-induced magnetic islands, along with comparisons with multiple numerical diagnostics. Our results demonstrate that the renormalization method, LE, and WBA consistently identify the field line onset stochasticity threshold. The critical magnetic island width predicted using this Hamiltonian method is significantly lower than the Chirikov criterion. 

Notably, the predicted critical island width appears to be in the same range with recent experimental observations on ASDEX Upgrade~\cite{Willensdorfer2024}, where RMP-induced islands of about $1-2~\mathrm{cm}$ are found in ELM suppression. However, the threshold for stochasticity onset due to the interaction between edge MHD islands is found to be significantly lower. This suggests that the interaction of edge MHD islands, if they pre-exist, can be a key mechanism for generating a stochastic layer, even before they interact with the RMP island.

Future work should focus on developing self-consistent Hamiltonian models that incorporate more accurate plasma response, and extending the present approach to the interactions among magnetic islands from beyond two rational surfaces.

%------------------------Acknowledgement-------------------------------
\section*{Acknowledgement}
This work was supported by the Undergraduate Training Program for Innovation, Entrepreneurship (Grant No. S202410487096), the National Magnetic Confinement Fusion Program of China (Grant No. 2019YFE03050004) and the Hubei International Science and Technology Cooperation Project under Grant No. 2022EHB003. The computing work in this paper is supported by the Public Service Platform of High Performance Computing by Network and Computing Center of HUST. We would also like to express our gratitude to Fangyuan Ma and Jiaxing Liu for their valuable help and discussions throughout this project.
%--------------------------------\appendix-----------------------------
\appendix
\section{Transformation to paradigm Hamiltonian}\label{APP.A}
This appendix starts from MHD equilibrium and perturbations, first constructs the Hamiltonian of magnetic field lines, expands and simplifies the Hamiltonian near the resonance surface, and finally obtains the paradigm Hamiltonian, and then relates it to the magnetic island width.
% ---------------------------------------------%
\subsection{Construct the Hamiltonian}
Ref.~[\onlinecite{Pina1988}] established the analogy of the action principle for magnetic field lines and of that for Hamiltonian mechanics~\cite{Escande2024} (for a more recent review, see~Ref.~[\onlinecite{RLVianaHamiltonian2023}]). Following these works, the Hamiltonian can be expressed as
\begin{equation}
H = - A_{\varphi},
\end{equation}
where $A_{\varphi}$ is the toroidal component of the vector potential. For the equilibrium magnetic field, we have $B_{0\theta}=-\partial A_\varphi/\partial r$. So we get 
\begin{equation}
H=-A_{\varphi}=\int \mathrm{d}r B_{0\theta}.
\end{equation}
Considering the safety factor 
\begin{equation}
q(r)=\frac{rB_{0\varphi}}{R_{0}B_{0\theta}(r)},
\end{equation}
the equilibrium Hamiltonian can be rewritten as
\begin{equation}
H=\int B_{0\theta} \mathrm{d}r=\frac{B_{0\varphi}}{R_{0}}\int\frac{r}{q(r)}\mathrm{d}r.
\end{equation}

We define the action variable as $J=r^2/2$, and the Hamiltonian can be expressed as
\begin{equation}
\begin{aligned}
H&=\frac{B_{0\varphi}}{R_{0}}\int\frac{r}{q(r)}\mathrm{d}r
=\frac{B_{0\varphi}}{R_{0}}\int\frac{1}{q(r)}r\mathrm{d}r\\
&=\frac{B_{0\varphi}}{R_{0}}\int\frac{1}{q(J)}\mathrm{d}J
\end{aligned}
\end{equation}
and rescale the new Hamiltonian $H_0=R_{0}H/B_{0\varphi}$~\cite{RLVianaHamiltonian2023}, we get
\begin{equation}
H_0=\int\frac{\mathrm{d}J}{q(J)}.
\end{equation}

Consider the generic perturbation field
\begin{equation}\begin{aligned}
\epsilon\mathbf{B}_1 
&=\epsilon\mathbf{B}_1^{(1)}(r)\exp\left[i(m_1\theta-n_{1}\varphi)\right] \\
&+\epsilon\mathbf{B}_1^{(2)}(r)\exp\left[i(m_2\theta-n_{2}\varphi)\right]
\end{aligned}
\end{equation}
and $\epsilon \ll 1$ marks the perturbation order and $\mathbf{B}_1(r)=(B_{1r}(r),B_{1\theta}(r),B_{1\varphi}(r))$. The perturbation Hamiltonian is
\begin{equation}
H_{1} = \frac{R_{0}}{B_{0\varphi}} 
\int \mathrm{d}r \epsilon B_{1\theta}(r).
\end{equation}
i.e. the complex perturbation Hamiltonian is
\begin{equation}
\tilde{H}_1^{(j)}(J,\theta,\varphi)
=-e^{i(m_{j}\theta-n_j\varphi)}\frac{R_{0}}{B_{0\varphi}} 
\int^J
\frac{\epsilon B_{1\theta}^{(j)}(\sqrt{2J^{\prime}})}{\sqrt{2J^{\prime}}}
\mathrm{d}J^{\prime},
\end{equation}
where j=1,2, denoting the mode number. Let the integral function 
\begin{equation}
I_j(J)\equiv 
\frac{R_{0}}{B_{0\varphi}} \int^J
\frac{B_{1\theta}^{(j)}(\sqrt{2J^{\prime}})}{\sqrt{2J^{\prime}}}
\mathrm{d}J^{\prime}.
\end{equation}
So the real perturbation coupled by two modes is
\begin{equation}
\begin{aligned}
H_1(J,\theta,\varphi)
&=-\epsilon \sum_{j=1}^2\Re\left\{I_j(J)e^{i(m_{j}\theta-n_j\varphi)}\right\}\\
&=-\epsilon \sum_{j=1}^2|I_j(J)|\cos(m_{j}\theta-n_j\varphi)
\end{aligned},
\end{equation}
where we choose the random phase of the complex function $I_{j}(J)$ to be zero. Finally, we get the total Hamiltonian
\begin{equation}
H(J,\theta,\varphi)=H_0(J)-\epsilon \sum_{j=1}^2|I_j(J)|\cos(m_{j}\theta-n_j\varphi)
\end{equation}
where
\begin{equation}
H_0(J)=\int\frac{\mathrm{d}J}{q(J)},\quad I_j(J)=\frac{R_{0}}{B_{0\varphi}} \int^J\frac{B_{1\theta}^{(j)}(\sqrt{2J^{\prime}})}{\sqrt{2J^{\prime}}}\mathrm{d}J^{\prime}.
\end{equation}
% ---------------------------------------------%
\subsection{The equation of motion and expansion}
We start from the Hamiltonian expressed in the new action-angle variables
\begin{equation}\label{eq:Hamiltonian-start}
H(J,\theta,\varphi) = H_0(J) + \epsilon \sum_{j=1}^{2} V_j(J) 
\cos\phi_j ,
\end{equation}
where the phase is defined as
\begin{equation}
\phi_j \equiv m_{j} \theta - n_j \varphi,
\label{eq:phase}
\end{equation}
and
\begin{equation}
H_0(J) = \int^J \frac{\mathrm{d}J'}{q(J')}, 
\qquad 
V_j(J) \equiv |I_j(J)| .
\label{eq:H0-Vj}
\end{equation}
We denote
\begin{equation}
\sigma(J) = H_0''(J), \qquad 
v_j = \frac{n_j}{m_{j}}.
\label{eq:sigma-v}
\end{equation}
The canonical equations of Hamiltonian~\eqref{eq:Hamiltonian-start} are
\begin{align}
\dot{\theta} &= \frac{\partial H}{\partial J} 
= H_0'(J) + \epsilon \sum_{j=1}^{2} V_j'(J) \cos\phi_j ,
\label{eq:eq-motion-theta} \\
\dot{J} &= - \frac{\partial H}{\partial \theta} 
= \epsilon \sum_{j=1}^{2}  m_{j} V_j(J) \sin\phi_j ,
\label{eq:eq-motion-J}
\end{align}
where the dot denotes $\mathrm{d}/\mathrm{d}\varphi$ since here $\varphi$ plays the role of a time variable.

Let $J=J_*+\delta J$ with $J_*$ a value at the reference rational surface. Expanding to first order, we have
\begin{equation}\label{eq:expand-H0} 
H_0'(J) \simeq w + \sigma_* \delta J , 
\end{equation}
where $w \equiv H_0'(J_*), \quad \sigma_* \equiv H_0''(J_*)$. And we also have
\begin{equation}\label{eq:expand-Vj}
V_j(J) \simeq V_{j*} + V'_{j*} \delta J,
\end{equation}
with $V_{j*}=V_j(J_*)$, $V'_{j*}=\mathrm{d}V_j/\mathrm{d}J|_{J_*}$. 
Substituting Eqs.~(\ref{eq:expand-H0}) and~(\ref{eq:expand-Vj}) into Eqs.~(\ref{eq:eq-motion-theta}) and~(\ref{eq:eq-motion-J}) gives
\begin{align}
\dot{\theta} &= w + \sigma_* \delta J 
+ \epsilon \sum_{j=1}^{2}  V'_{j*}\cos\phi_j , 
\label{eq:theta-expand} \\
\delta\dot{J} &= \epsilon \sum_{j=1}^{2} m_{j}\left(V_{j*}+V'_{j*}\delta J\right)\sin\phi_j .
\label{eq:J-expand}
\end{align}
Differentiating \cref{eq:theta-expand} with respect to $\varphi$ yields
\begin{equation}
\ddot{\theta} = \sigma_* \delta\dot{J} 
+ \epsilon \sum_{j=1}^{2}  V'_{j*}(-\sin\phi_j)\dot{\phi}_j .
\label{eq:theta-second}
\end{equation}
By Eqs.~(\ref{eq:phase}), (\ref{eq:sigma-v}) and~(\ref{eq:theta-expand}), we have
\begin{equation}\label{eq:phi-dot}
\begin{aligned}
\dot{\phi}_j 
&= m_{j} \dot{\theta} - n_j\\
&= m_{j} (w-v_j) + m_{j}\sigma_* \delta J 
+ m_{j} \epsilon \sum_\ell V'_{\ell*}\cos\phi_\ell ,
\end{aligned}
\end{equation}
Using \cref{eq:J-expand}, we obtain after algebra
\begin{equation}\label{eq:theta-second-final}
\begin{aligned}
\ddot{\theta} 
&= \epsilon \sum_{j} m_{j} \Phi_j \sin\phi_j \\
&- \frac{\epsilon^2}{2}\sum_{j,\ell} m_{j} V'_{j*}V'_{\ell*}
\Big[\sin(\phi_j+\phi_\ell)+\sin(\phi_j-\phi_\ell)\Big],
\end{aligned}
\end{equation}
where the effective amplitude is defined as
\begin{equation}
\Phi_j \equiv \sigma_* V_{j*} + V'_{j*}(v_j-w).
\label{eq:Phi-def}
\end{equation}
Importantly, the $\delta J$--dependent terms cancel exactly, so that only the phases $\phi_j$ remain. Note that this reduction recovers the centered-resonance approximation introduced by Chirikov (see Sec.3.1.5 of~Ref.~[\onlinecite{Escande1985}]), where one expands about the resonance center and retains the leading symmetric terms, which justifies neglecting higher-order asymmetric corrections in the present derivation.
% ---------------------------------------------%
\subsection{The paradigm Hamiltonian}
\cref{eq:theta-second-final} is equivalent to the one--dimensional time--dependent Hamiltonian
\begin{equation}
H'(P,\theta,\varphi) = \frac{P^2}{2} 
+ \epsilon \sum_{j=1}^{2} \Phi_j \cos\phi_j(\theta,\varphi) 
+ \mathcal{O}(\epsilon^2),
\label{eq:Hprime}
\end{equation}
with canonical pair $(\theta,P)$ and $\dot{\theta}=P$. Introducing the phase
\begin{equation}
Y \equiv m_1 \theta - n_1 \varphi - \frac{\pi}{2}.
\label{eq:Y-def}
\end{equation}
Here $Y$ is a slow phase because we expand about the reference rational surface $J_*$ where the resonance condition $m_j w\approx n_j$ holds, thus $\dot\phi_j=m_j(w-v_j)+\mathcal{O}(\sigma_*\delta J,\epsilon)$ is small.
And rescaling time with the phase--velocity difference
\begin{equation}
\Delta v \equiv |v_2-v_1| , \qquad \tau = \Delta v \, \varphi ,
\label{eq:delta-v}
\end{equation}
the momentum variable can be rescaled to
\begin{equation}
X \equiv \frac{P}{\Delta v},
\end{equation}
so that the time-dependent phase entering the second wave is
\begin{equation}
Z(\varphi)=m_1\,\tau-\frac{\pi}{2}=m_1\Delta v\,\varphi-\frac{\pi}{2}.
\end{equation}
As a result, the Hamiltonian can be transformed to the paradigm form
\begin{equation}\label{eq:Hpar}
H_{\mathrm{par}}(X,Y,Z) = \tfrac{1}{2}X^2 
- M \cos Y - P \cos\!\big[k(Y-Z)\big],
\end{equation}
where $k=m_2/m_1$, and
\begin{equation}
M = \frac{\epsilon \,\Phi_1}{\Delta v^2},
\qquad 
P = \frac{\epsilon \,\Phi_2}{\Delta v^2}.
\label{eq:M-P}
\end{equation}

At leading order in the slow--variation approximation ($V'_{j*}\sim\mathcal{O}(\eta)\ll1$), 
the effective amplitudes reduce to
\begin{equation}
\Phi_j \simeq \sigma_* V_{j*},
\label{eq:Phi-leading}
\end{equation}
so that the two normalized resonance magnitudes are proportional to the local curvature $\sigma_*=H_0''(J_*)$ and the mode amplitudes $V_{j*}$.
% ---------------------------------------------%
\subsection{Connection with the island width}
We now relate the normalized amplitudes $M,P$ to the magnetic island widths $W_j$.
Recall Zohm's formula (see Eq.~(8.55) of~Ref.~[\onlinecite{Zohm2014-yc}]) for the island width (evaluated at the rational surface $r_s$):
\begin{equation}\label{eq:magnetic_island_width_repeat}
W_j = 4 \sqrt{\frac{B_{1r}^{(j)}(r_s)\; r_s \; q(r_s)}{m_{j} \; q'(r_s)\; B_{0\theta}(r_s)}},
\end{equation}
which implies
\begin{equation}
B_{1r}^{(j)}(r_s)=\frac{m_{j} \, q'(r_s)\, B_{0\theta}(r_s)}{16\, r_s \, q(r_s)} \; W_j^2 .
\label{eq:Br_from_W}
\end{equation}

Use the definition
$$
V_j(J) = |I_j(J)|, \qquad 
I_j(J)=\frac{R_{0}}{B_{0\varphi}} \int^{J}\frac{B_{1\theta}^{(j)}(\sqrt{2J'})}{\sqrt{2J'}}\,\mathrm{d}J',
\label{eq:Vj_Ij_def}
$$
one readily checks (change variable $J'=r'^2/2$, $\mathrm{d}J' = r'\mathrm{d}r'$) that at the rational surface $J_{*}=r_s^2/2$
\begin{equation}
I_j(J_*)=\frac{R_{0}}{B_{0\varphi}} \int_{0}^{r_s} B_{1\theta}^{(j)}(r')\,\mathrm{d}r' .
\label{eq:Ij_radial_int}
\end{equation}
Using the solenoidal condition for a single Fourier harmonic $ \propto e^{i(m_{j}\theta-n_j\varphi)} $ and neglecting the small toroidal perturbation component, one obtains (in cylindrical approximation)
\begin{equation}
\partial_r\big(r B_{1r}^{(j)}\big) + i m_{j} B_{1\theta}^{(j)} = 0
\end{equation}
which yields
\begin{equation}\label{eq:div_free_relation}
r_s B_{1r}^{(j)}(r_s) = - i m_{j} \int_{0}^{r_s} B_{1\theta}^{(j)}(r')\,\mathrm{d}r' = - i m_{j} \frac{B_{0\varphi}}{R_{0}} I_{j}(J_{*}).
\end{equation}
Taking magnitudes and using \cref{eq:Ij_radial_int} gives the useful estimate
\begin{equation}\label{eq:Vj_from_Br}
V_{j*} \equiv |I_j(J_*)| \simeq \frac{R_{0}}{B_{0\varphi}} \frac{r_s\,|B_{1r}^{(j)}(r_s)|}{m_{j}} .
\end{equation}
Combining Eqs.~(\ref{eq:Br_from_W}) and~(\ref{eq:Vj_from_Br}) yields
\begin{equation}\label{eq:Vj_in_terms_of_W}
V_{j*} \simeq \frac{R_{0}}{B_{0\varphi}} \frac{q'(r_s)\,B_{0\theta}(r_s)}{16\,q(r_s)} \; W_j^2 .
\end{equation}

Next, recall the effective amplitude $\Phi_j$ used in the slow-variation reduction, we get
\begin{equation}\label{eq:Phi_sigma}
\Phi_j \simeq \sigma_*\, V_{j*}
\end{equation}
and
\begin{equation} 
\sigma_* \equiv H_0''(J_*) = -\frac{\mathrm d q/\mathrm d J}{q^2}\Big|_{J_*}.
\end{equation}
Since $\mathrm{d}q/\mathrm{d}J = 1/r~\mathrm{d}q/\mathrm{d}r$, we may write
\begin{equation}
\sigma_* = -\frac{q'(r_s)}{r_s \, q(r_s)^2} .
\label{eq:sigma_in_r}
\end{equation}
Substituting Eqs.~(\ref{eq:Vj_in_terms_of_W}) and~(\ref{eq:sigma_in_r}) into~\cref{eq:Phi_sigma} gives
\begin{equation}
\Phi_j \simeq -\,\frac{R_{0}}{B_{0\varphi}} \,\frac{q'(r_s)^2\, B_{0\theta}(r_s)}{16\, r_s \, q(r_s)^3}\; W_j^2 .
\label{eq:Phi_in_terms_of_W_intermediate}
\end{equation}

Finally, expressing $B_{0\theta}(r_s)$ via the safety factor $q(r_s)=r_s B_{0\varphi}/(R_0 B_{0\theta}(r_s))$ (i.e.\ $B_{0\theta}(r_s)=r_s B_{0\varphi}/(R_0 q(r_s))$), \cref{eq:Phi_in_terms_of_W_intermediate} can be simplified to
\begin{equation}\label{eq:Phi_in_terms_of_W}
\Phi_j 
\simeq -\,\frac{R_{0}}{B_{0\varphi}} \,\frac{q'(r_s)^2\, B_{0\varphi}}{16\, R_{0} \, q(r_s)^4}\; W_j^2 
= -\,\frac{q'(r_s)^{2}}{16\, q(r_s)^{4}}\; W_j^2
\end{equation}
Using the relation between $\Phi_j$ and the normalized amplitudes (cf.\ \cref{eq:M-P}):
\begin{equation}\label{eq:M-P_repeat}
M = \frac{\epsilon\,\Phi_1}{\Delta v^2}, \qquad
P = \frac{\epsilon\,\Phi_2}{\Delta v^2},
\end{equation}
we obtain the desired representation of $M,P$ in terms of island widths
\begin{equation}
\begin{aligned}
\label{eq:M-P_in_terms_of_W}
M \simeq \epsilon\,\frac{q'(r_s)^{2}}{16\, q(r_s)^{4}}\; 
\frac{W_1^2}{\left(\frac{n_{2}}{m_{2}}-\frac{n_{1}}{m_{1}}\right)^2},
\\
P \simeq \epsilon\,\frac{q'(r_s)^{2}}{16\, q(r_s)^{4}}\; 
\frac{W_2^2}{\left(\frac{n_{2}}{m_{2}}-\frac{n_{1}}{m_{1}}\right)^2}.
\end{aligned}
\end{equation}
Here we displayed the positive amplitude; the sign of $\Phi_j$ may be absorbed into the phase of the cosine potential.

\makeatletter
\renewcommand{\thesection}{\@arabic\c@section}
\makeatother
\renewcommand{\thefigure}{\arabic{figure}}

\nocite{*}
\bibliographystyle{aipnum4-1}
\bibliography{aipsamp}

% --------------- RMP_MHD_Function_Figures ---------------
\clearpage
\begin{figure}[htbp]
\centering
\includegraphics[width=\linewidth]{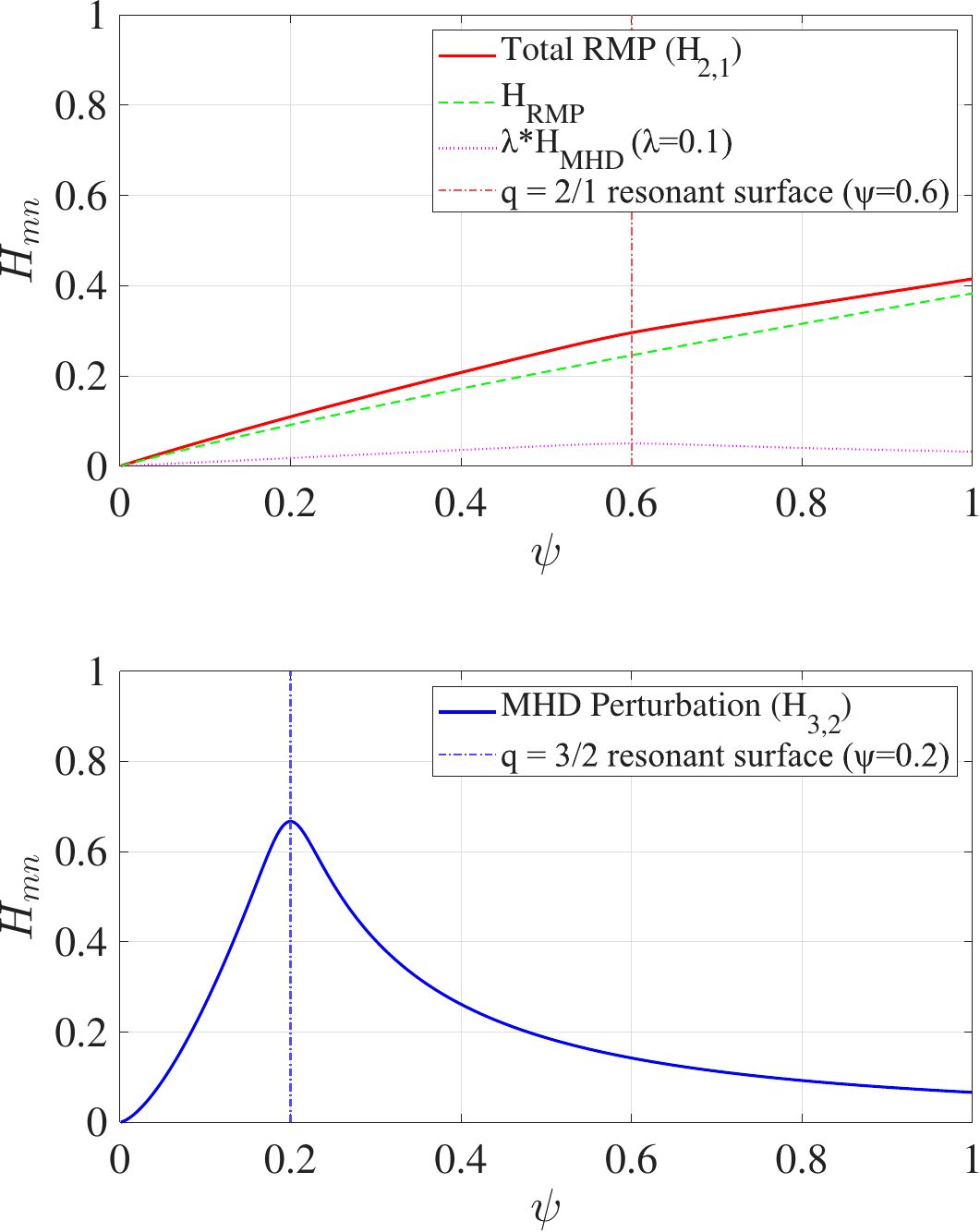}
\caption{Radial profiles of Hamiltonian perturbations $H_{mn}$ as a function of normalized toroidal flux $\psi$, calculated using the simulation parameters described in the text. Top: (2,1) perturbation, including RMP and MHD components; Bottom: (3,2) MHD perturbation.}
\label{fig:RMP perturbation and MHD perturbation diagrams}
\end{figure}

% --------------- q-profile figure ---------------
\clearpage
\begin{figure}[htbp]
\centering
\includegraphics[width=\linewidth]{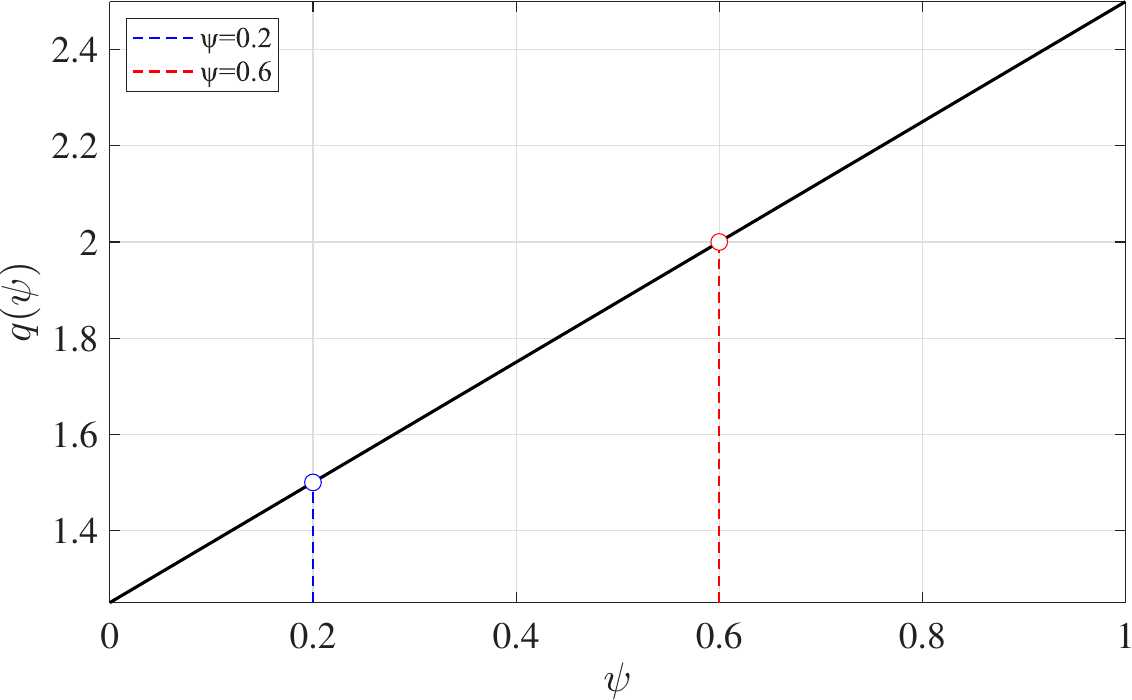}
\caption{q-profile as function of the normalized toroidal flux $\psi$.}
\label{fig:q-profile}
\end{figure}

% --------------- renormalization figure ---------------
\clearpage
\begin{figure}[htbp]
\centering
\includegraphics[width=\linewidth]{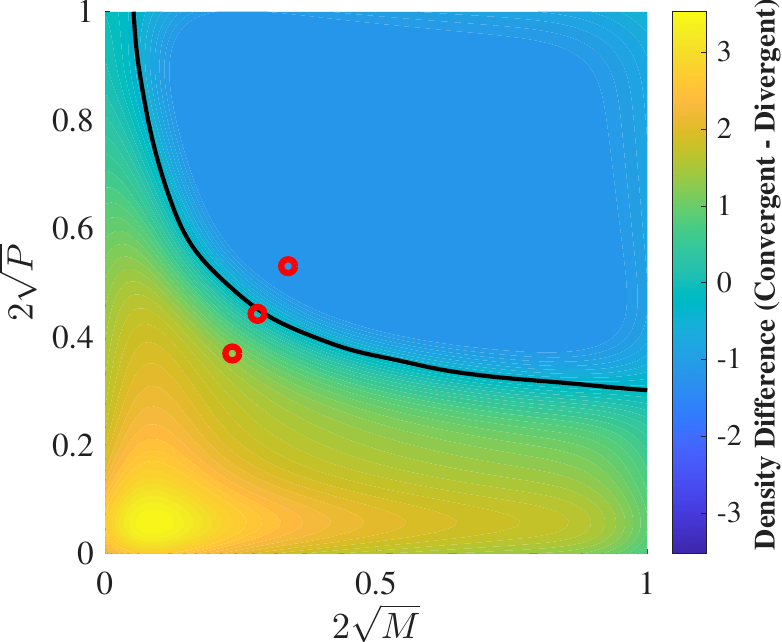}
\caption{
SVM and theoretical Boundary in $(2\sqrt{M}, 2\sqrt{P})$ space, separating the stable region (yellow) and unstable region (blue) with sample points for $\epsilon = 0.010, 0.014, 0.020$. Large scale stochasticity is predicted to occur as $\epsilon$ crosses the boundary.
}
\label{fig:renormalization_combined}
\end{figure}

% --------------- Poincaré maps figure --------------- 
\clearpage
\begin{figure}[htbp]
\centering
\begin{subfigure}[b]{\linewidth}
\centering
\includegraphics[width=0.8\linewidth]{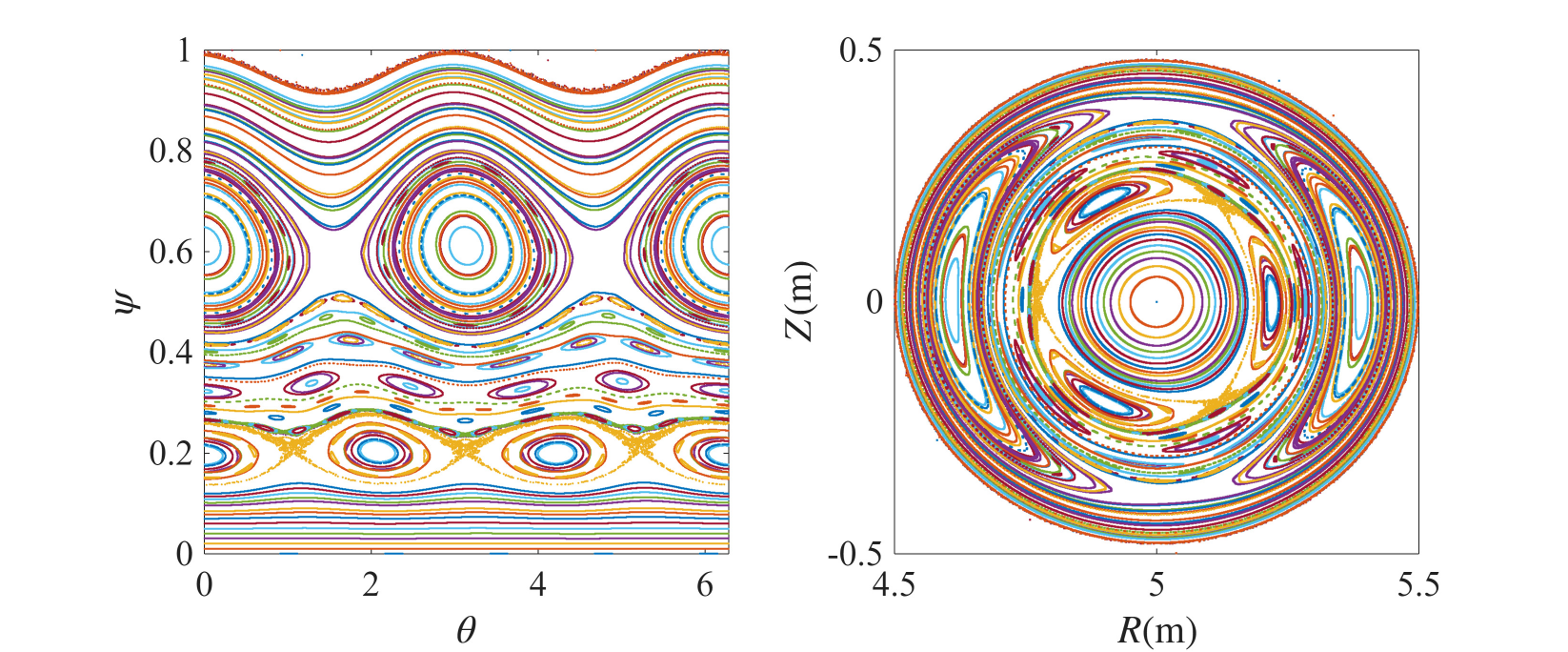}
\caption{$\epsilon=0.010$}
\end{subfigure}

\begin{subfigure}[b]{\linewidth}
\centering
\includegraphics[width=0.8\linewidth]{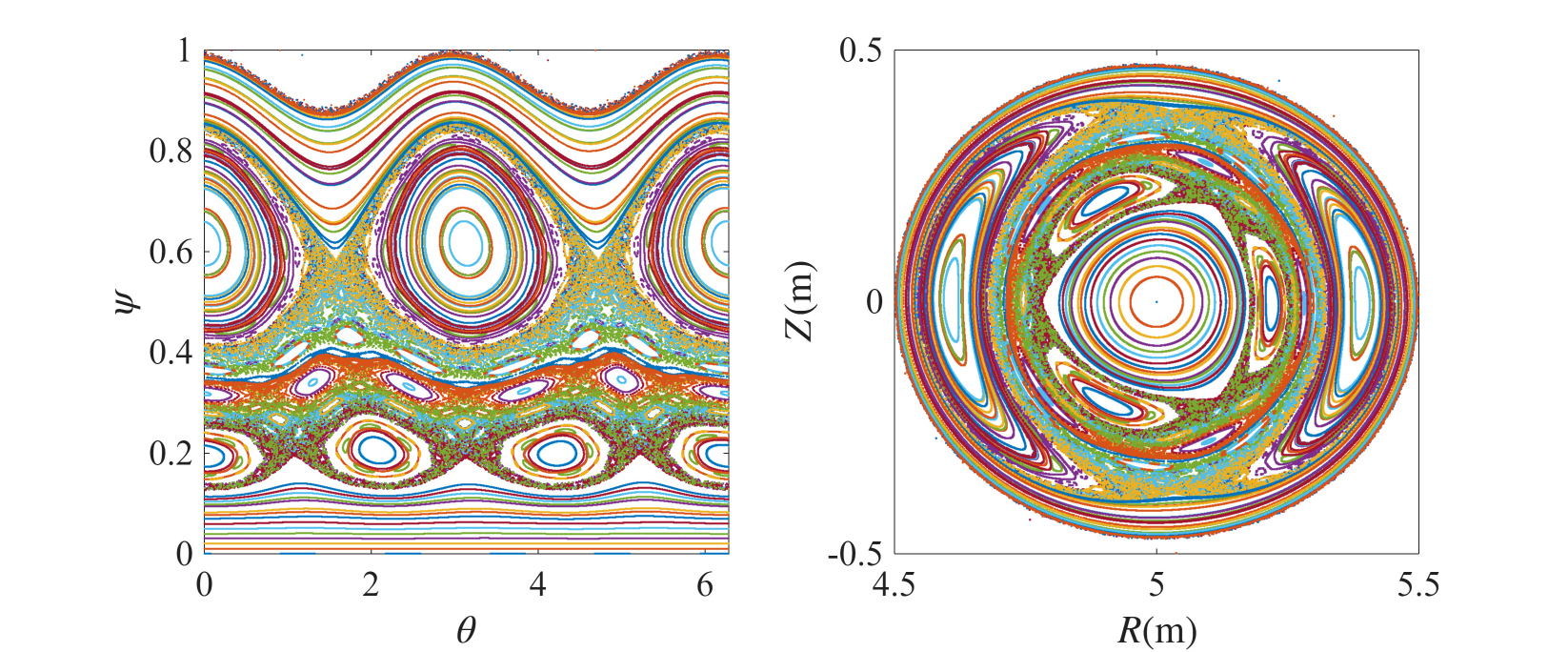}
\caption{$\epsilon=0.014$}
\end{subfigure}

\begin{subfigure}[b]{\linewidth}
\centering
\includegraphics[width=0.8\linewidth]{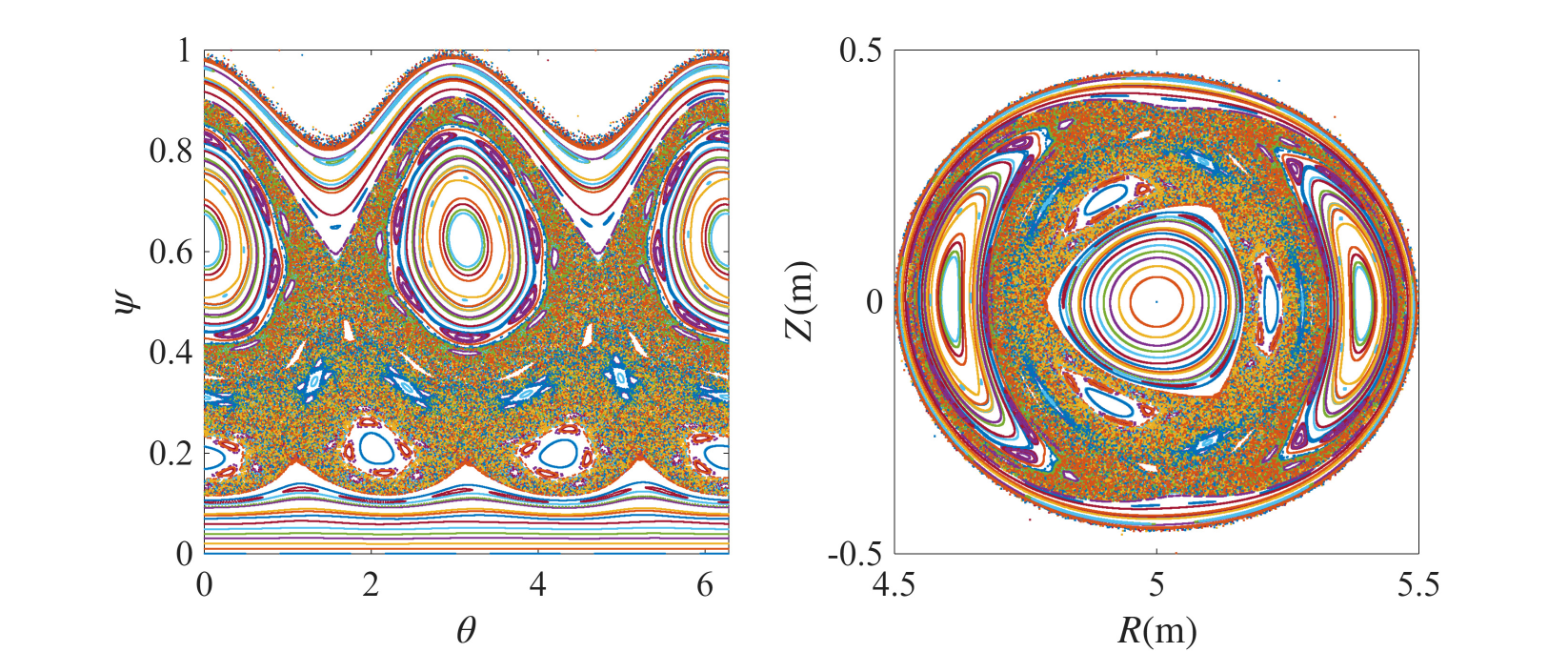}
\caption{$\epsilon=0.020$}
\end{subfigure}
\caption{Comparison of Poincaré maps for different perturbation magnitudes $\epsilon$. (a) $\epsilon=0.010$, (b) $\epsilon=0.014$, (c) $\epsilon=0.020$.}
\label{fig:poincare_maps}
\end{figure}

% --------------- Lyapunov exponent figure ---------------
\clearpage
\begin{figure}[htbp]
\centering
\begin{subfigure}[b]{\linewidth}
\centering
\includegraphics[width=0.64\linewidth]{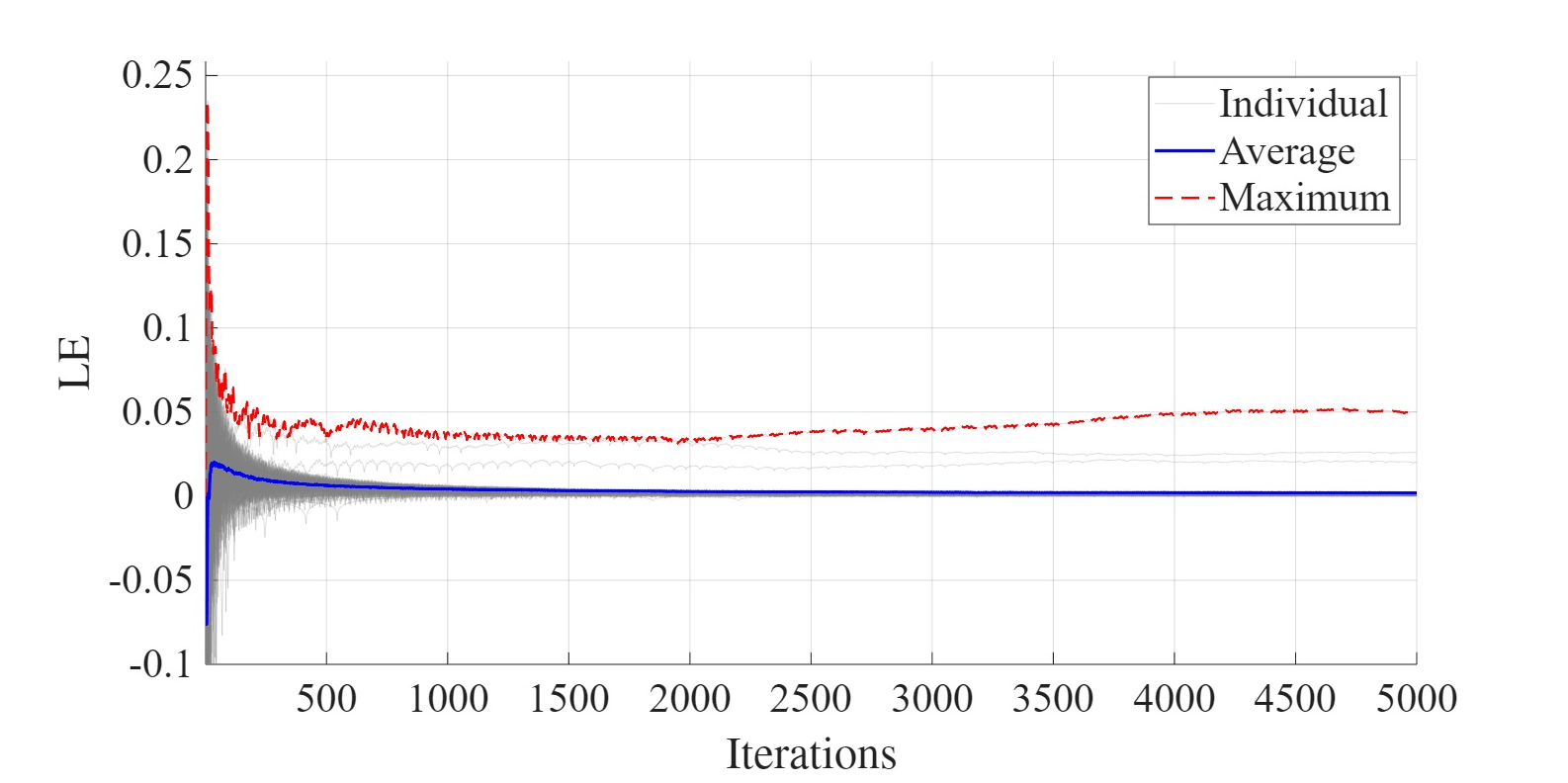}
\caption{$\epsilon=0.010$}
\end{subfigure}
\hfill
\begin{subfigure}[b]{\linewidth}
\centering
\includegraphics[width=0.64\linewidth]{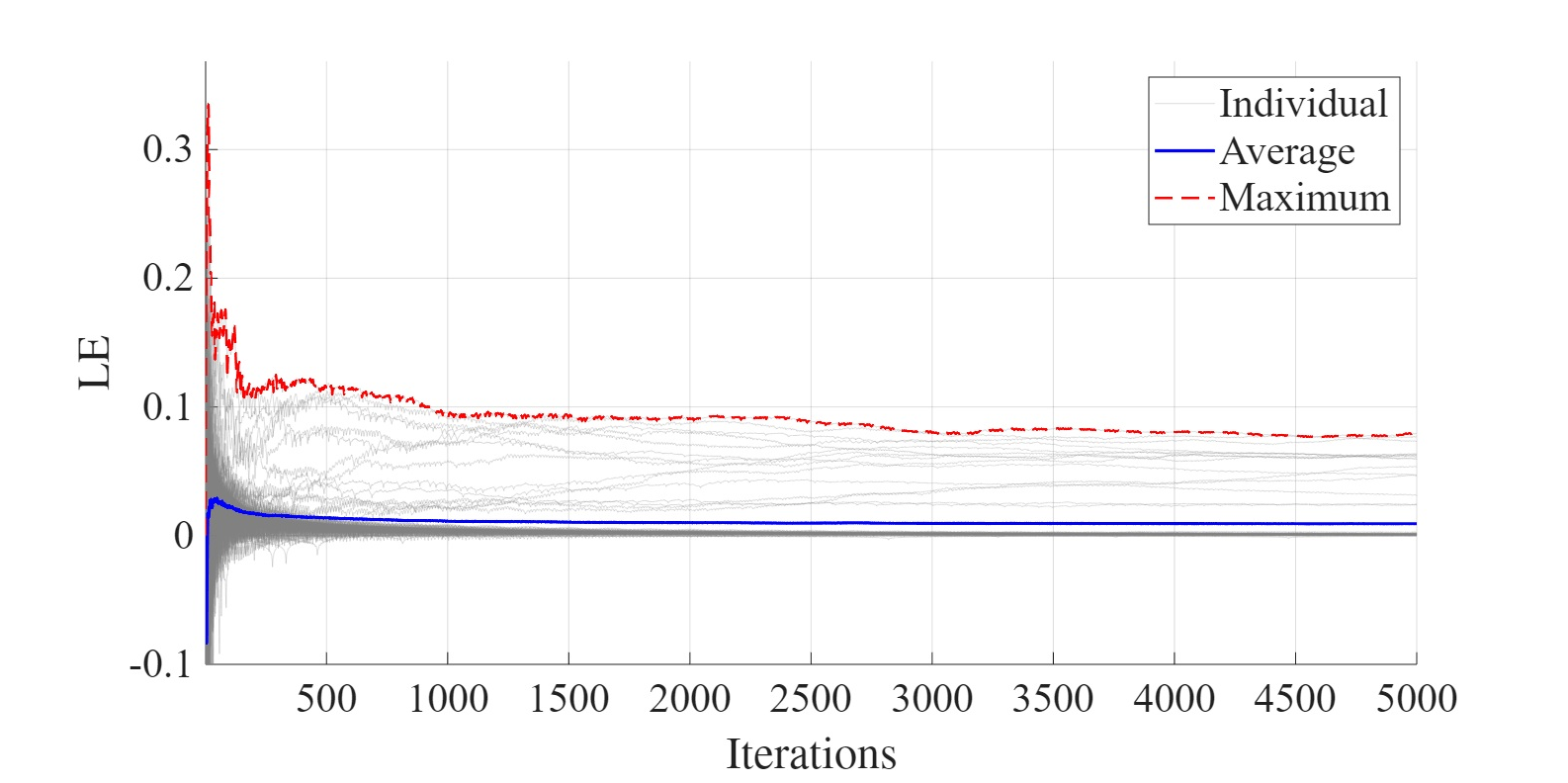}
\caption{$\epsilon=0.014$}
\end{subfigure}
\hfill
\begin{subfigure}[b]{\linewidth}
\centering
\includegraphics[width=0.64\linewidth]{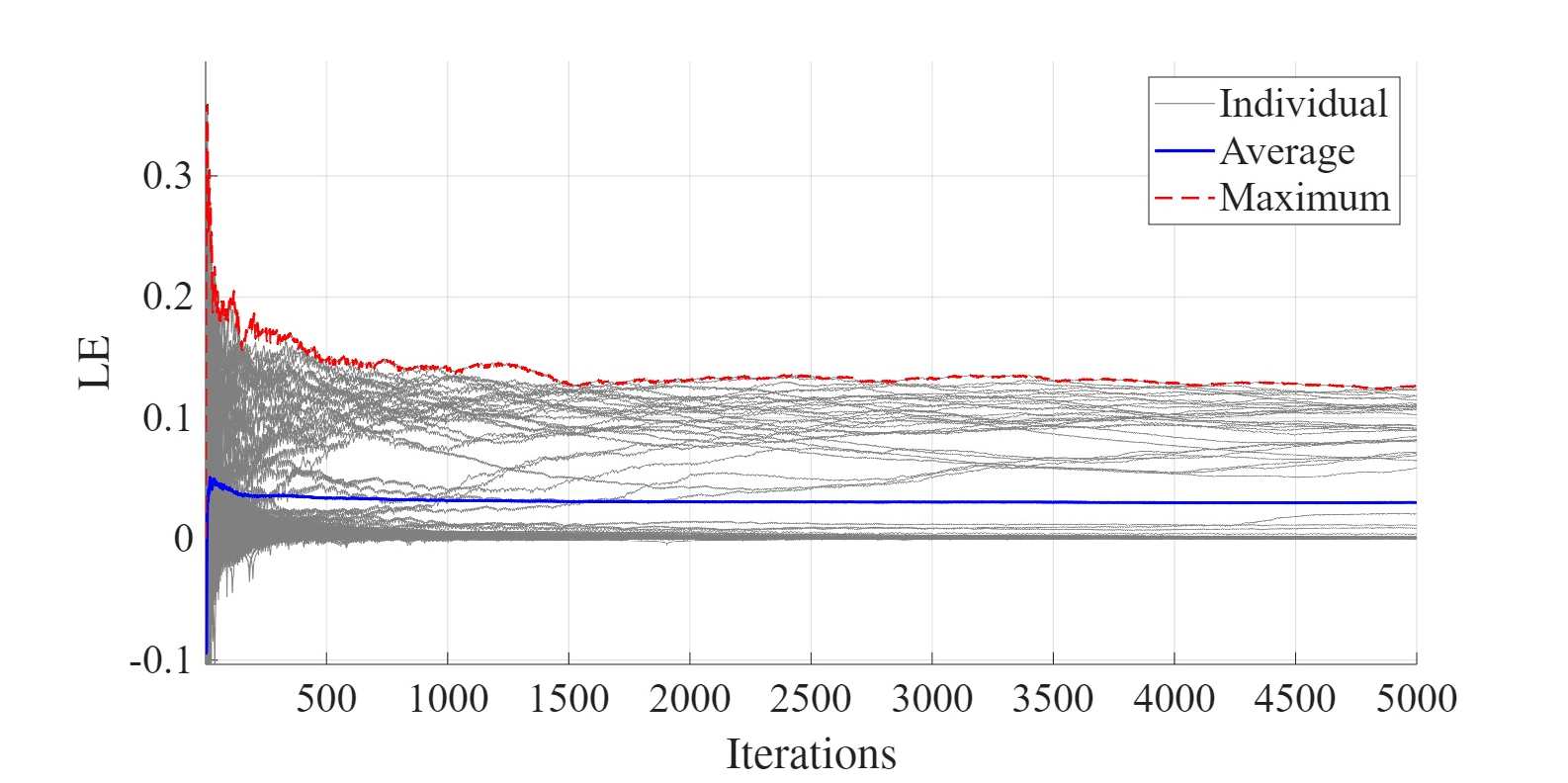}
\caption{$\epsilon=0.020$}
\end{subfigure}
\caption{Evolution of Lyapunov exponents for different perturbation magnitudes $\epsilon$. (a) $\epsilon=0.010$, (b) $\epsilon=0.014$, (c) $\epsilon=0.020$. The gray curve clusters illustrate the evolution process of the LEs $\lambda(N)$ for each magnetic field line trajectory between $\psi = 0$ and $\psi = 1$ surface as a function of the number of iterations $N$. The maximum LE, $\lambda_{\max}(N)$, is indicated by a red dashed line, while the system-averaged LE, $\langle\lambda(N)\rangle$, is represented by a blue solid line.}
\label{fig:le_evolution}
\end{figure}

% --------------- Results of weighted Birkhoff average -------------
\clearpage
\begin{figure}[htbp] 
\centering

% Epsilon = 0.010 %
\begin{subfigure}[b]{0.6\linewidth}
\centering
\includegraphics[width=\linewidth]{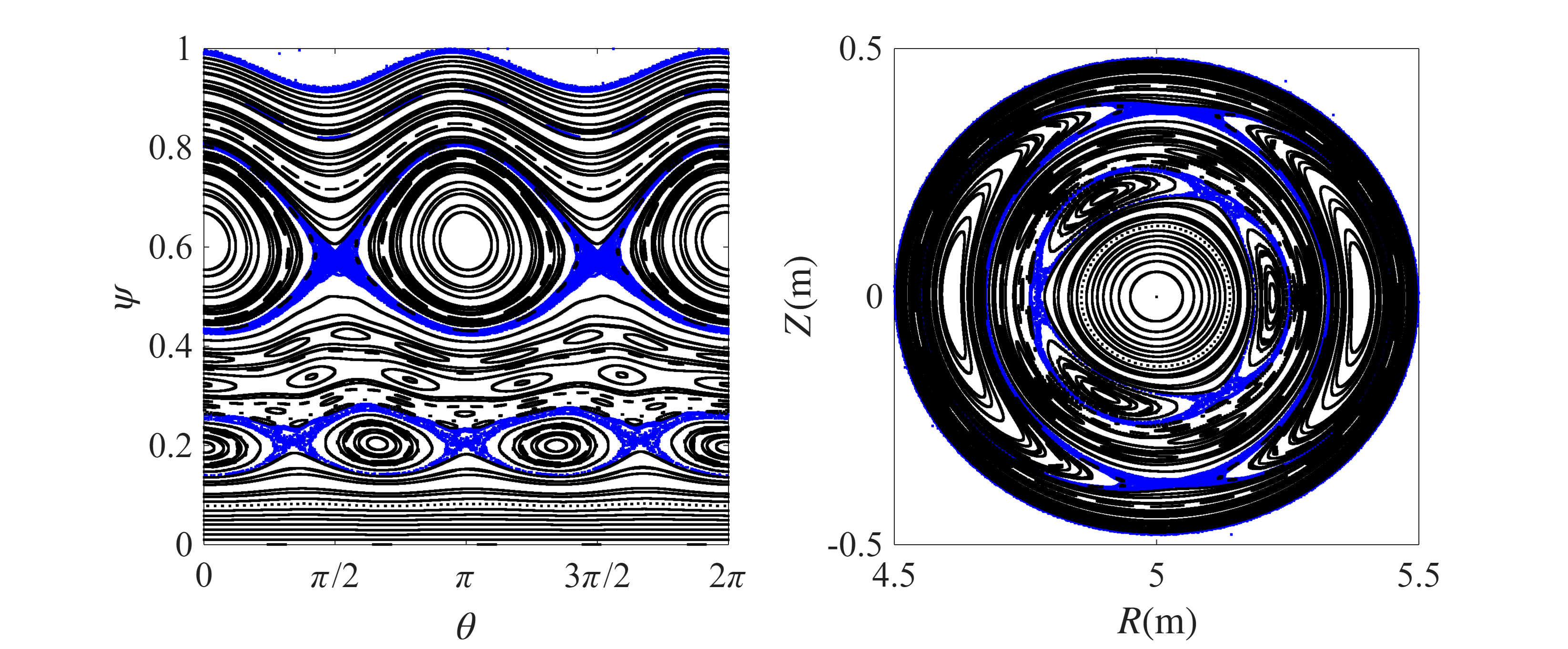}
\caption{Poincaré plot, $\epsilon=0.010$}
\end{subfigure}
\hfill
\begin{subfigure}[b]{0.35\linewidth}
\centering
\includegraphics[width=\linewidth]{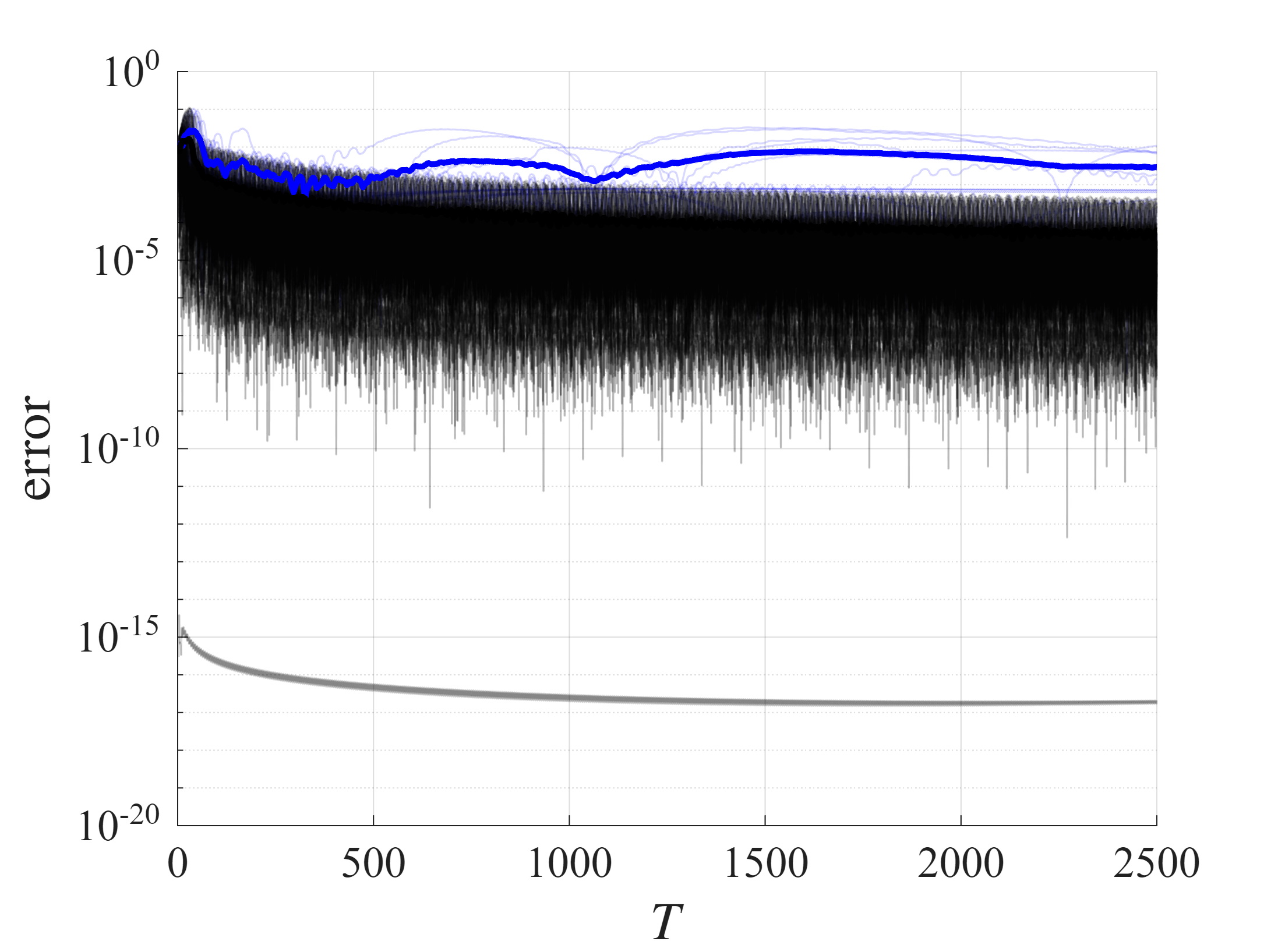}
\caption{WBA residuals, $\epsilon=0.010$}
\end{subfigure}

\vspace{1em} % Add some vertical space between rows

% Epsilon = 0.020 %
\begin{subfigure}[b]{0.6\linewidth}
\centering
\includegraphics[width=\linewidth]{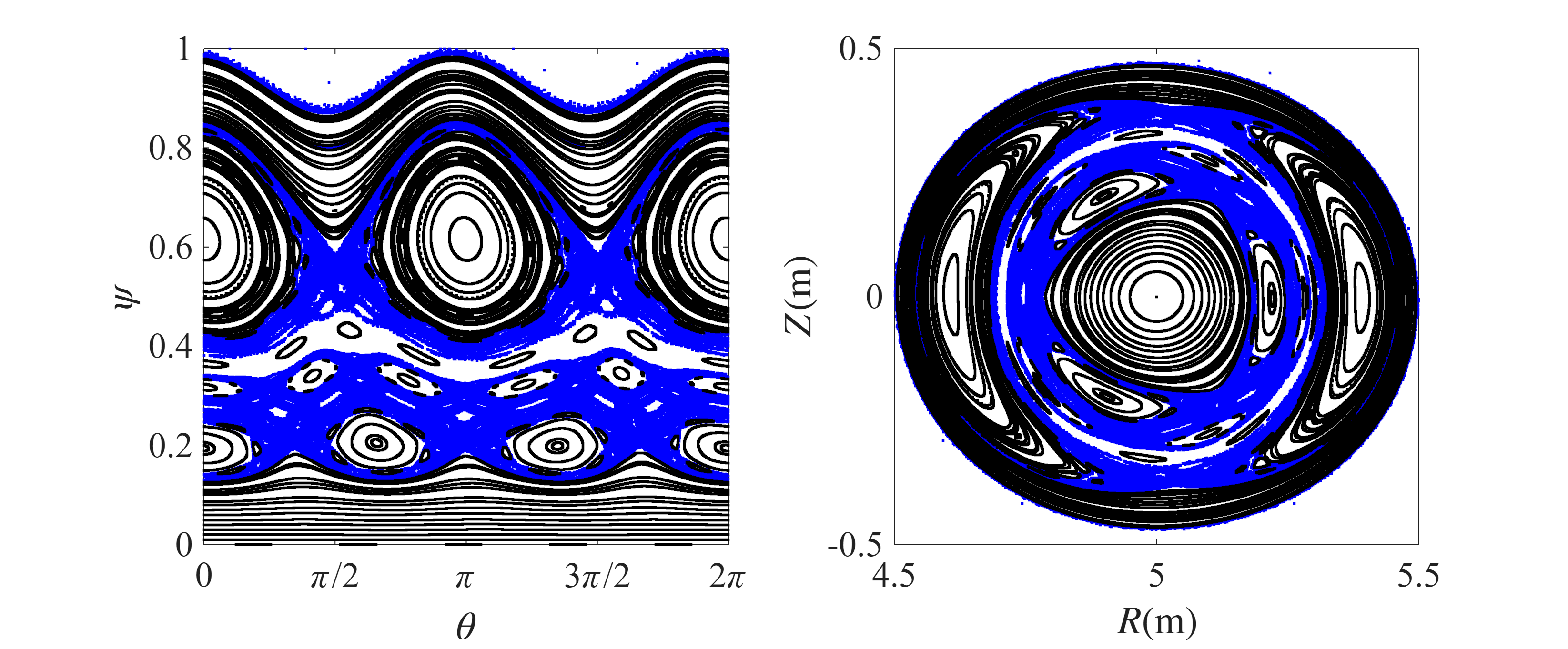}
\caption{Poincaré plot, $\epsilon=0.014$}
\end{subfigure}
\hfill
\begin{subfigure}[b]{0.35\linewidth}
\centering
\includegraphics[width=\linewidth]{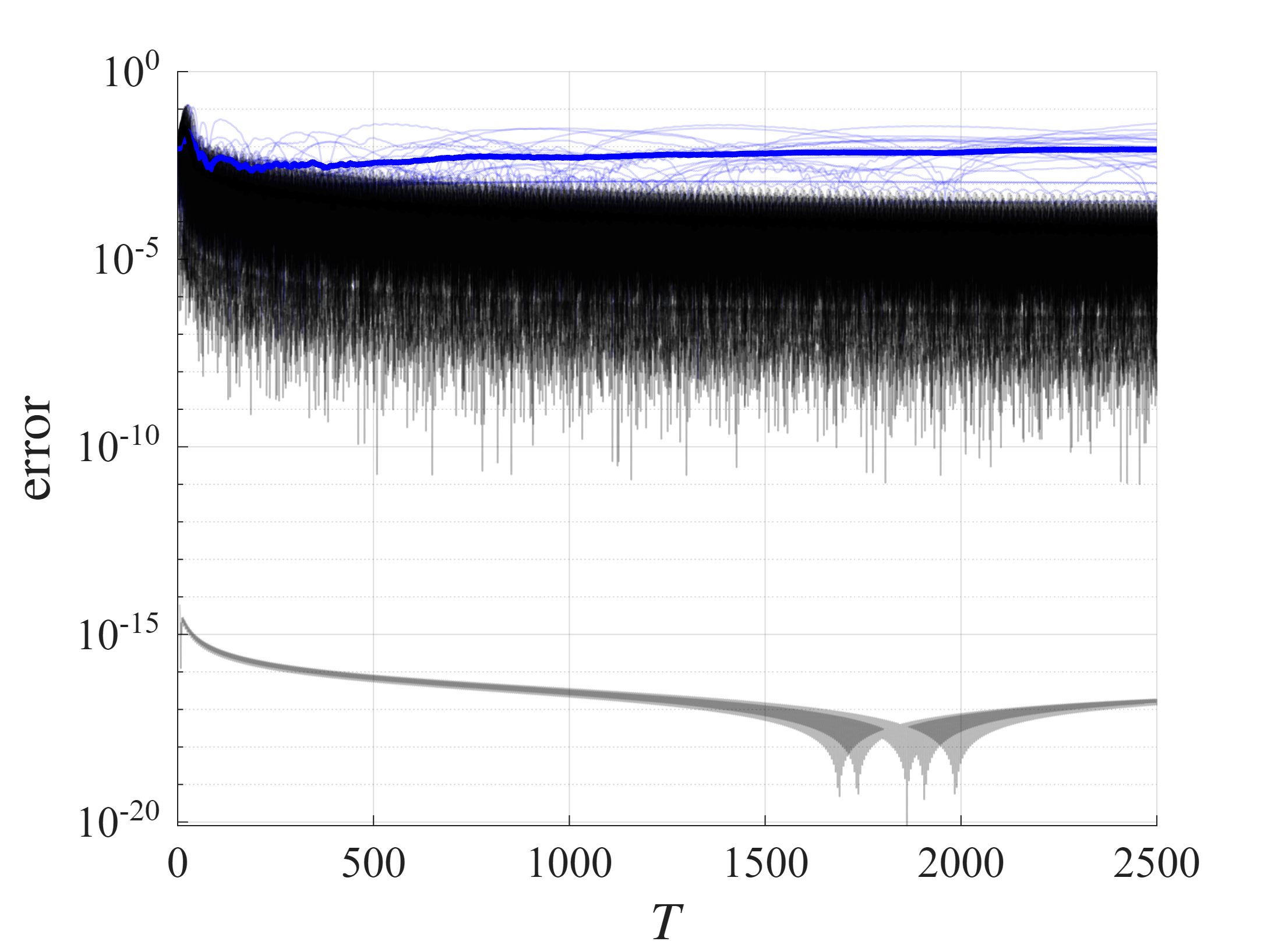}
\caption{WBA residuals, $\epsilon=0.014$}
\end{subfigure}

\vspace{1em} % Add some vertical space between rows

% Epsilon = 0.020 %
\begin{subfigure}[b]{0.6\linewidth}
\centering
\includegraphics[width=\linewidth]{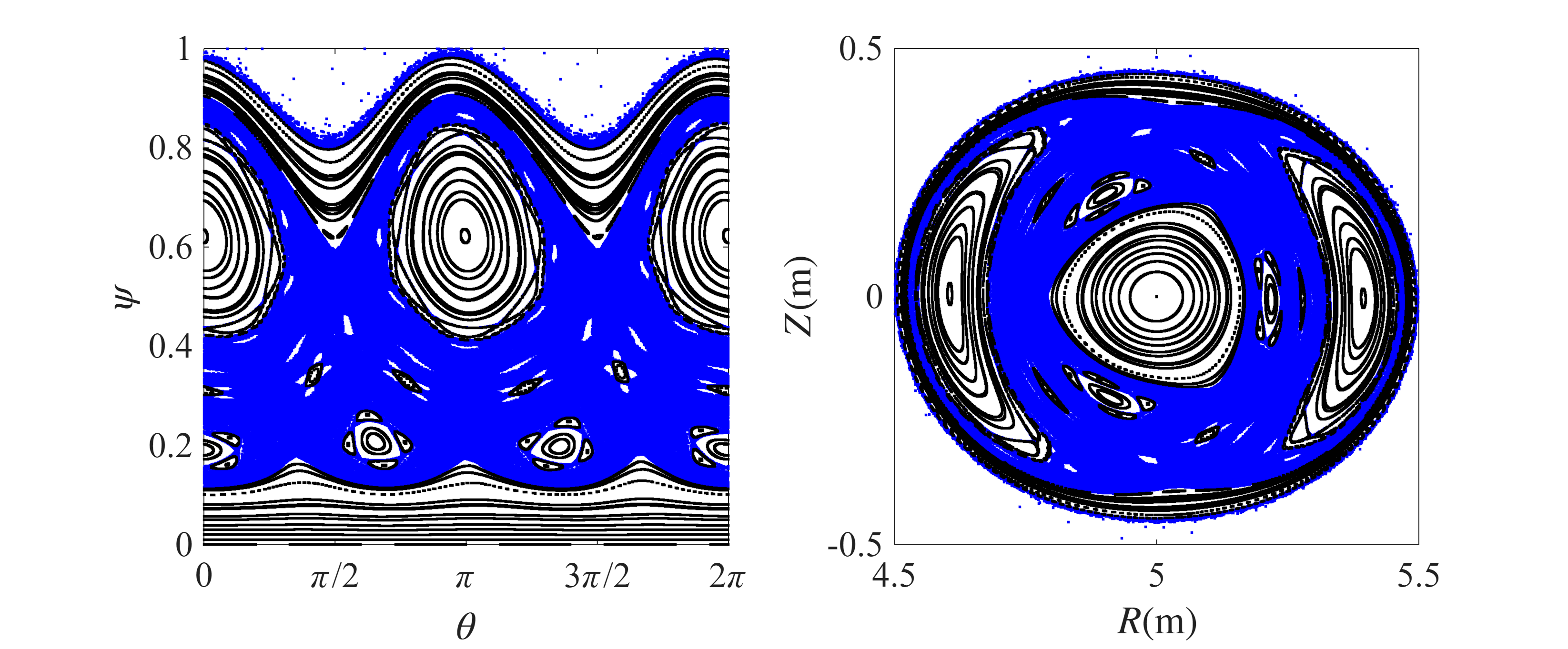}
\caption{Poincaré plot, $\epsilon=0.020$}
\end{subfigure}
\hfill
\begin{subfigure}[b]{0.35\linewidth}
\centering
\includegraphics[width=\linewidth]{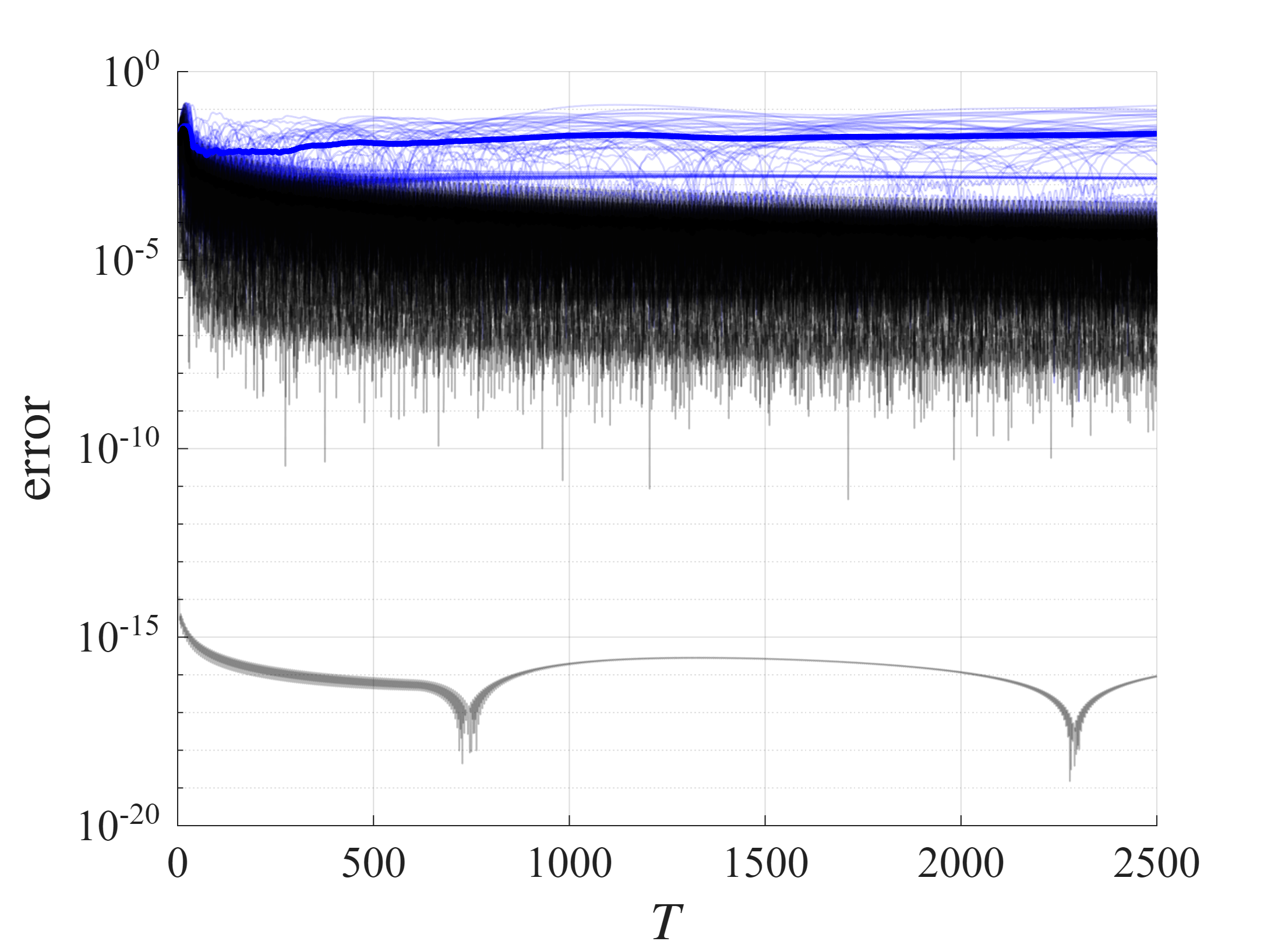}
\caption{WBA residuals, $\epsilon=0.020$}
\end{subfigure}
\caption{Left column (a, c, e): Poincaré plots with regular orbits (black) and chaotic orbits (blue). Right column (b, d, f): Convergence of WBA residuals, showing super-exponential convergence for regular orbits and slow convergence for chaotic ones.}
\label{fig:wba_analysis_combined}
\end{figure}

% --------------- Correlation figure ---------------
\clearpage
\begin{figure}[htbp]
\centering
%%%%
\begin{subfigure}[b]{\linewidth}
\centering
\includegraphics[width=0.65\linewidth]{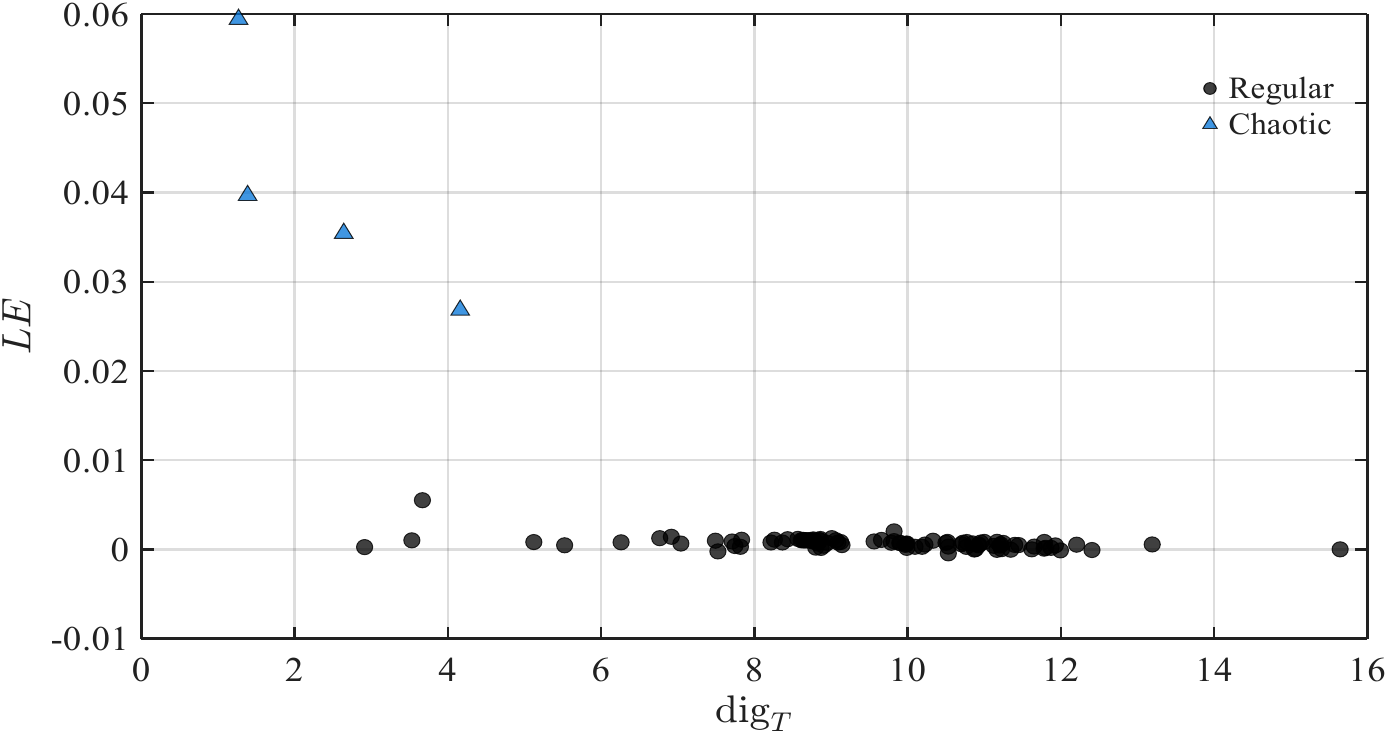}
\caption{$\epsilon=0.010$}
\end{subfigure}
\hfill
\begin{subfigure}[b]{\linewidth}
\centering
\includegraphics[width=0.65\linewidth]{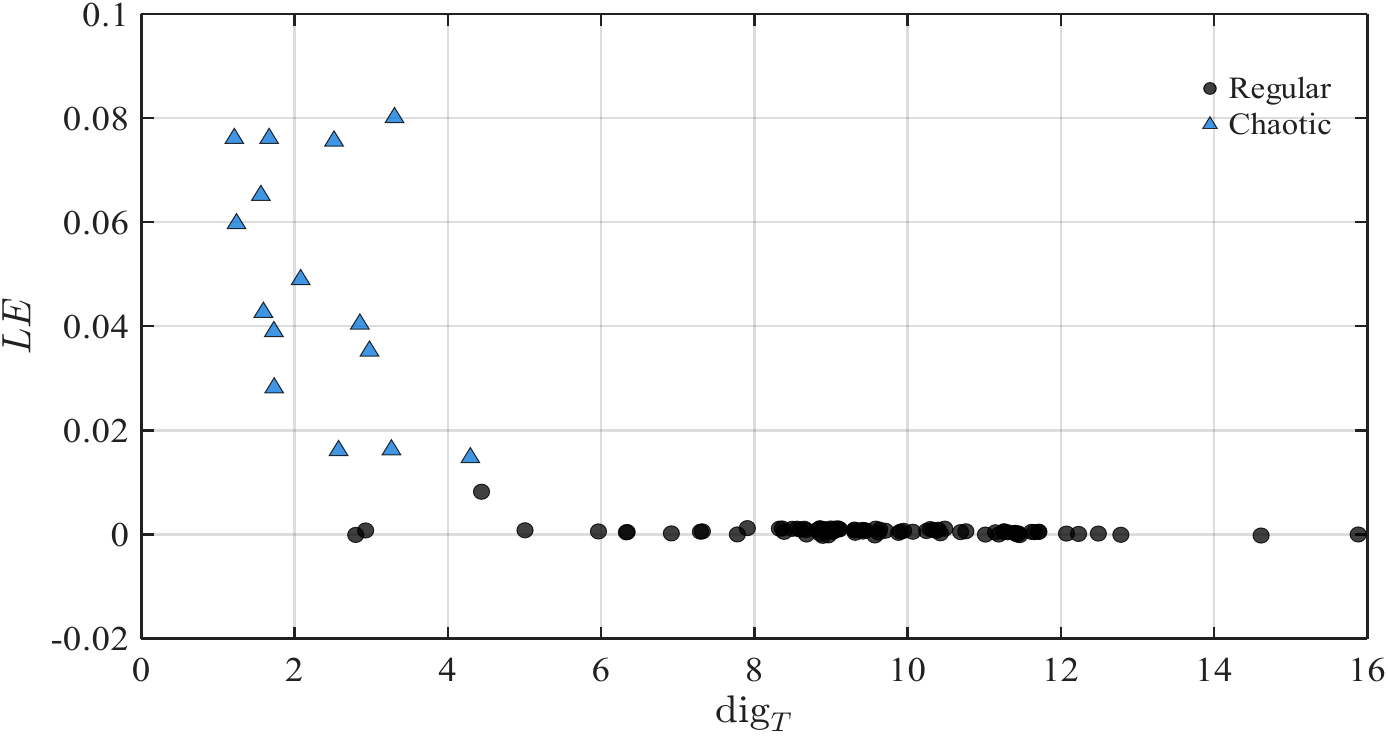}
\caption{$\epsilon=0.014$}
\end{subfigure}
\hfill
\begin{subfigure}[b]{\linewidth}
\centering
\includegraphics[width=0.65\linewidth]{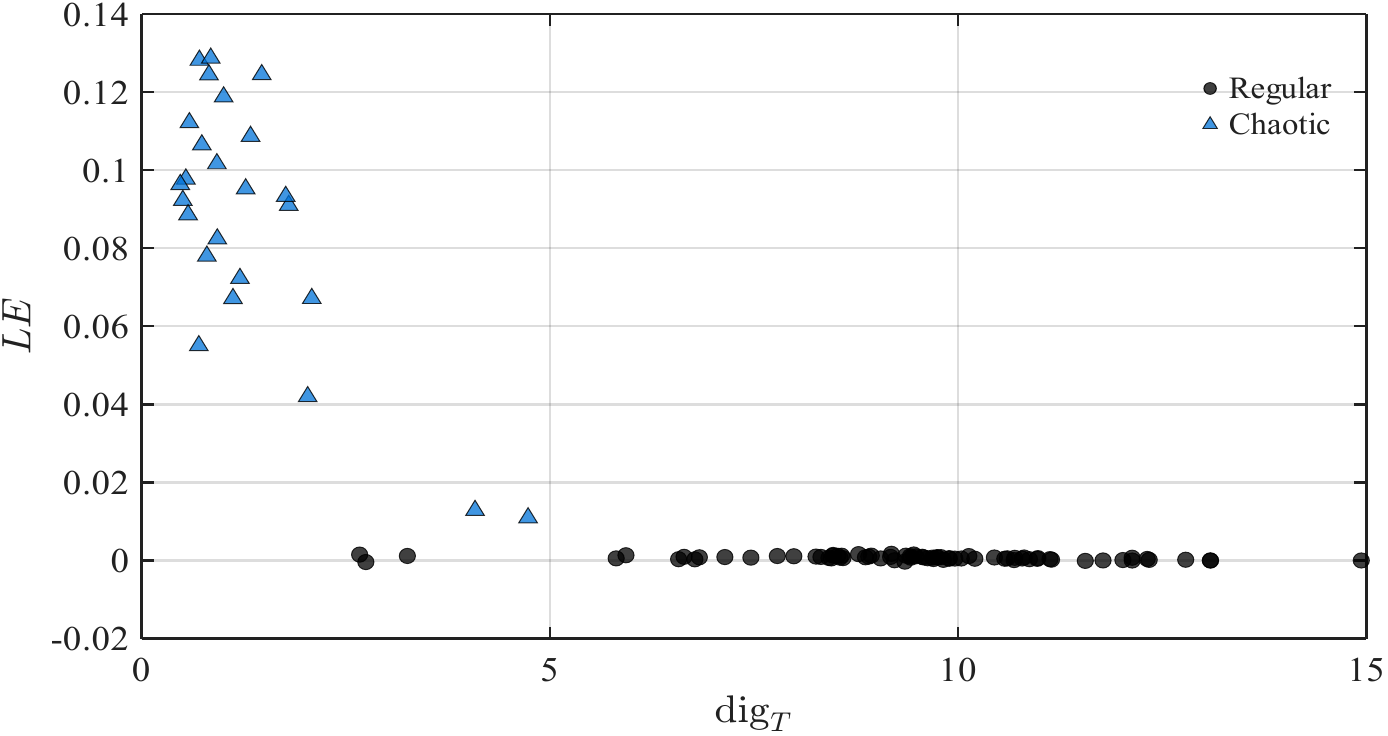}
\caption{$\epsilon=0.020$}
\end{subfigure}
%%%%
\caption{Correlation between the final LE and $\mathrm{dig}_T$ for different perturbation amplitudes: (a)$\epsilon = 0.010$, (b) $\epsilon = 0.014$, and (c) $\epsilon = 0.020$. Orbits are classified as regular (black triangles, LE $\approx 0$) or chaotic (blue circles, LE $> 0.01$).}
\label{fig:digT_vs_LE}
\end{figure}

% --------------- Golden Mean figure ---------------
\clearpage
\begin{figure}[htbp]
\centering
    \begin{subfigure}[b]{\linewidth}
    \includegraphics[width=0.8\linewidth]{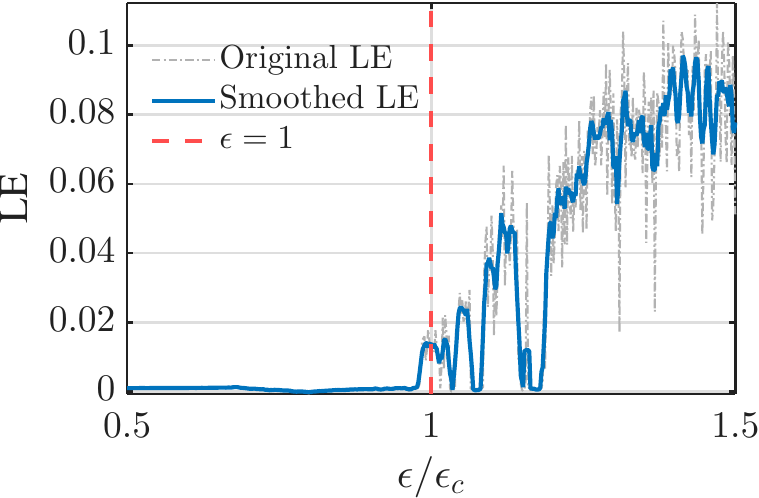}
    \label{fig:Golden Mean LE}
    \end{subfigure}
\hfill
    \begin{subfigure}[b]{\linewidth}
    \includegraphics[width=0.8\linewidth]{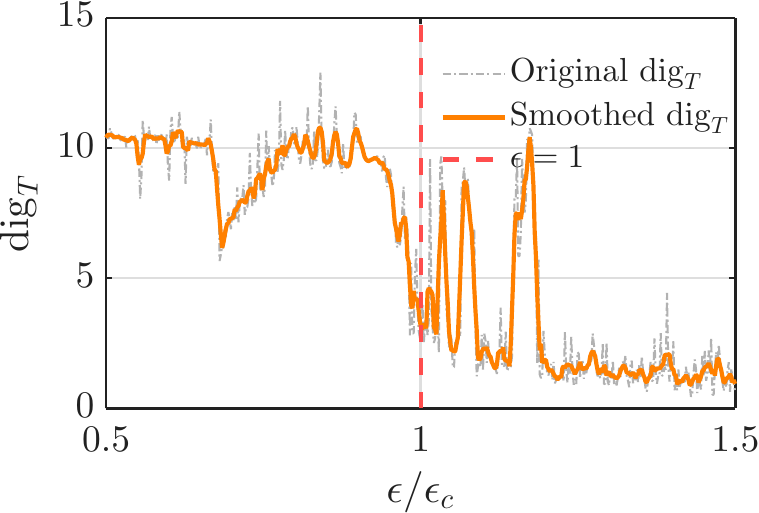}
    \label{fig:Golden Mean Dig_T}
    \end{subfigure}
\caption{
Upper: LE of the golden mean torus, with the dashed line at $\lambda = 10^{-2}$ indicating the chaos criterion. 
Down: Weighted Birkhoff average (WBA) digit accuracy ($\mathrm{dig}_T$) for the same torus, where the sharp drop signals torus breakup.
}
\label{fig:golden_mean_thresholds}
\end{figure}

% --------------- q_DIII-D_figure  --------------- 
\clearpage
\begin{figure}[htbp]
\centering
\includegraphics[width=\linewidth]{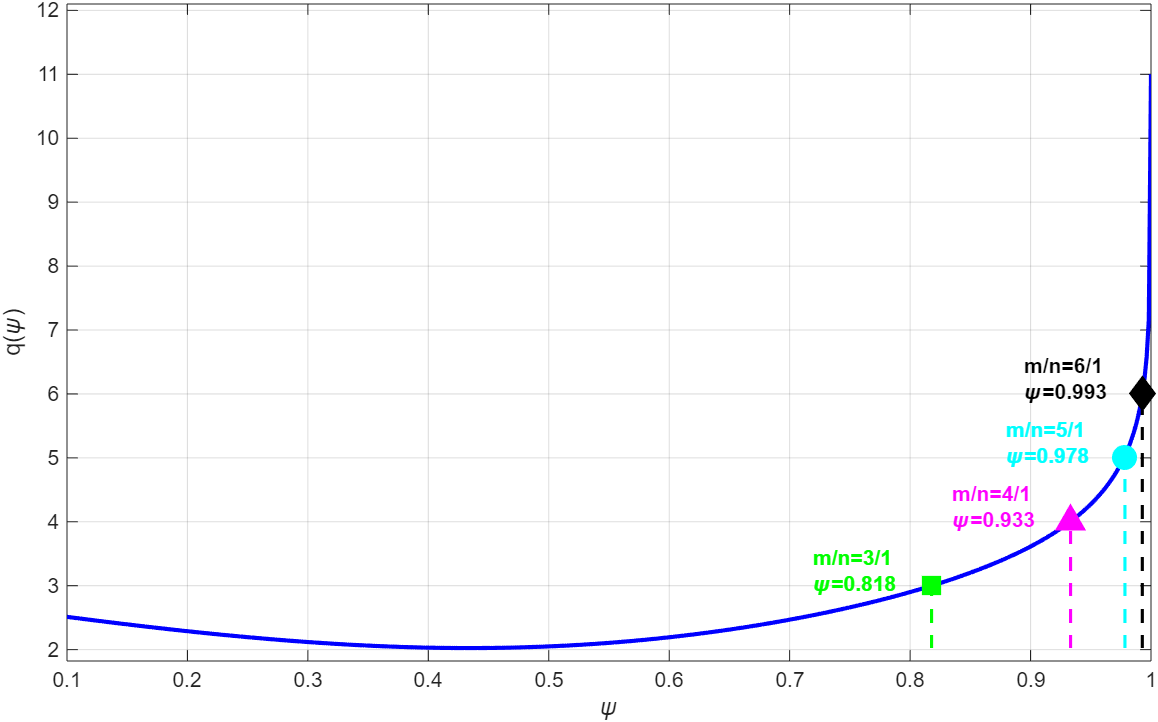}
\caption{Safety factor profile for DIII-D (blue line). Distinct rational surfaces $(m,n)$ are highlighted with different colors, with only a representative subset of magnetic surfaces explicitly marked for clarity.}
\label{fig:q_factor_DIII-D}
\end{figure}

% --------------- DIII-D_renormal_island_scatter ---------------
\clearpage
\begin{figure}[htbp]
\centering
    \begin{subfigure}[b]{\linewidth}
    \includegraphics[width=0.9\linewidth]{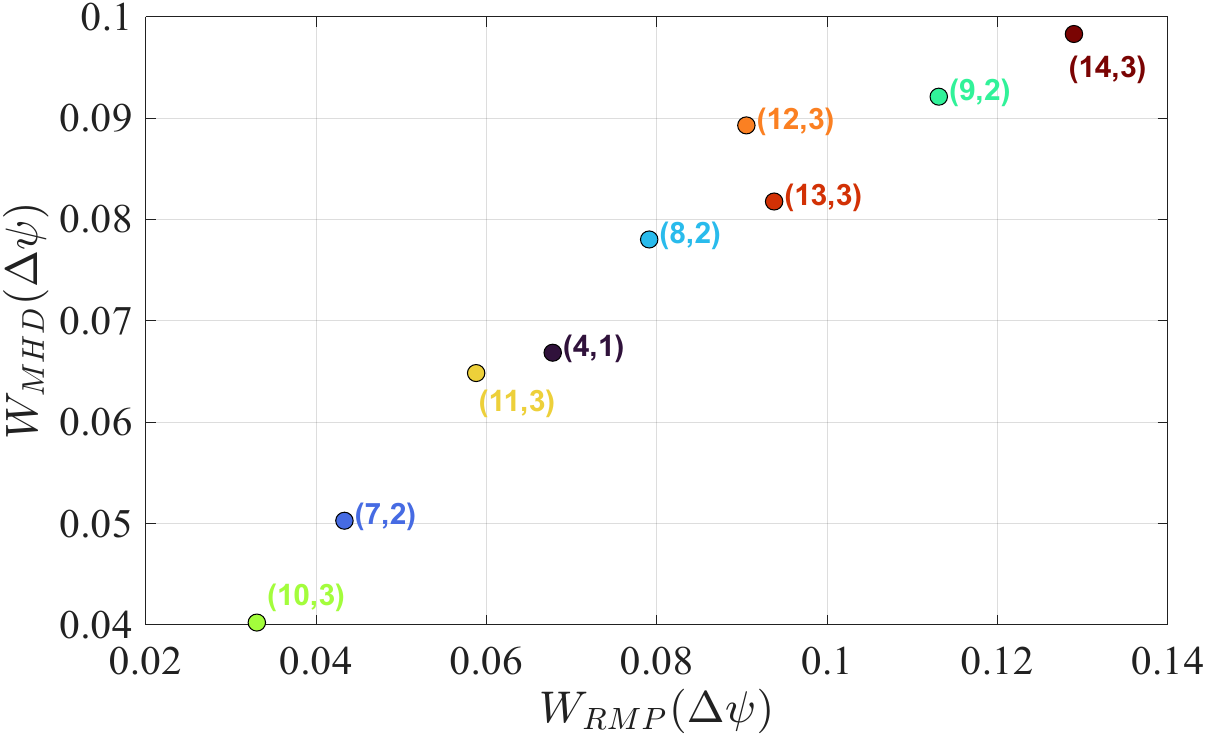}
    \end{subfigure}
\hfill
    \begin{subfigure}[b]{\linewidth}
    \includegraphics[width=0.9\linewidth]{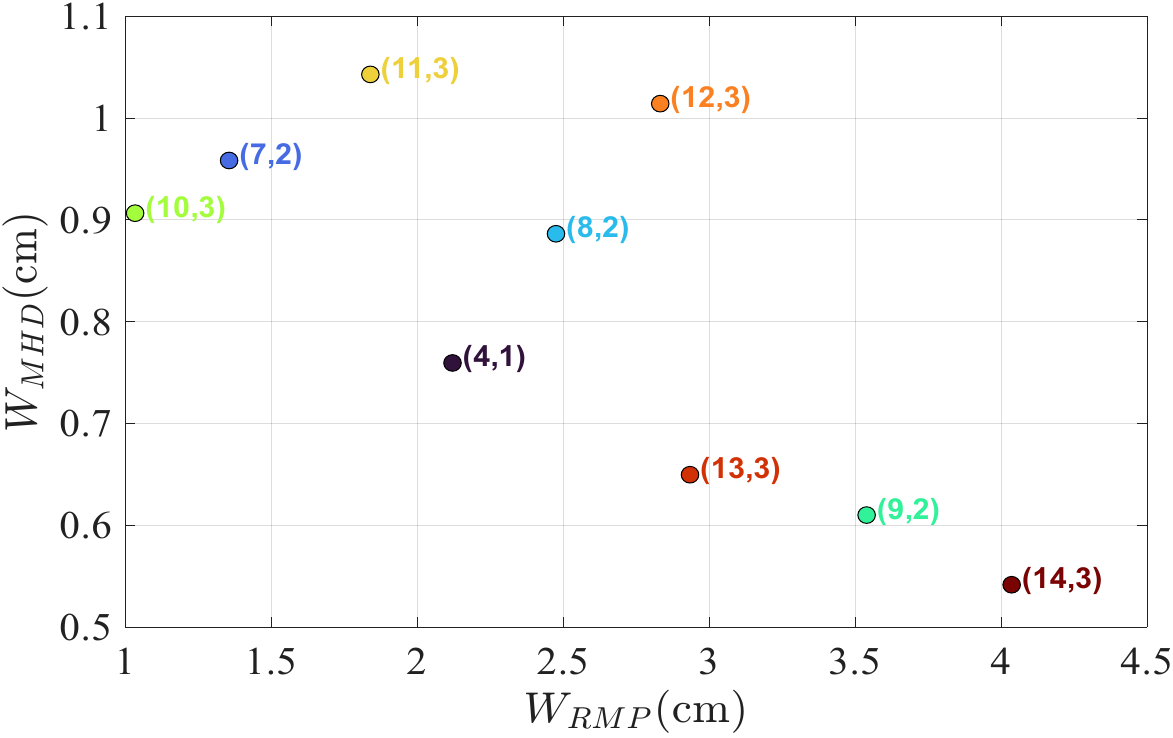}
    \end{subfigure}
\caption{
Critical Magnetic Island Width: RMP vs MHD. The upper subgraph shows the width relationship in the flux coordinate. The down subgraph shows the same relationship but is converted to a physical unit ($\mathrm{cm}$). Each point in the scatter plot represents a rational surface denoted by $(m,n)$.
}
\label{fig:RMPvsMHD}
\end{figure}

% --------------- DIIID-D_single_digT ---------------
\clearpage
\begin{figure}[htbp]
\centering
\includegraphics[width=0.95\linewidth]{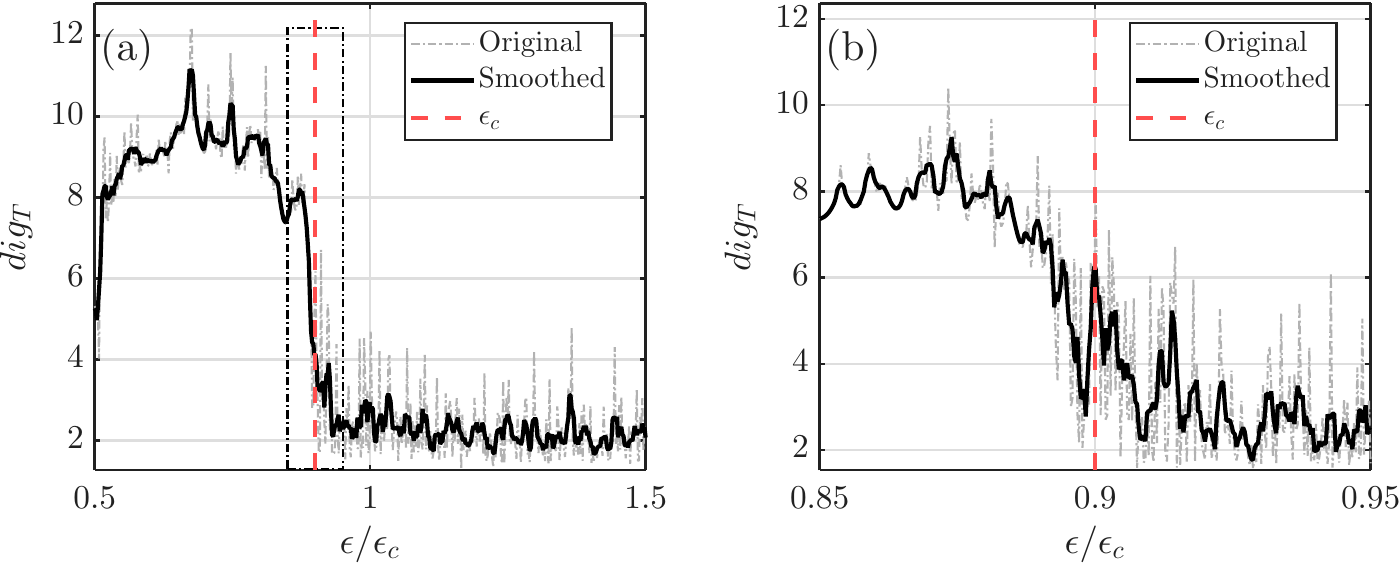}
\vspace{1em}
\includegraphics[width=0.95\linewidth]{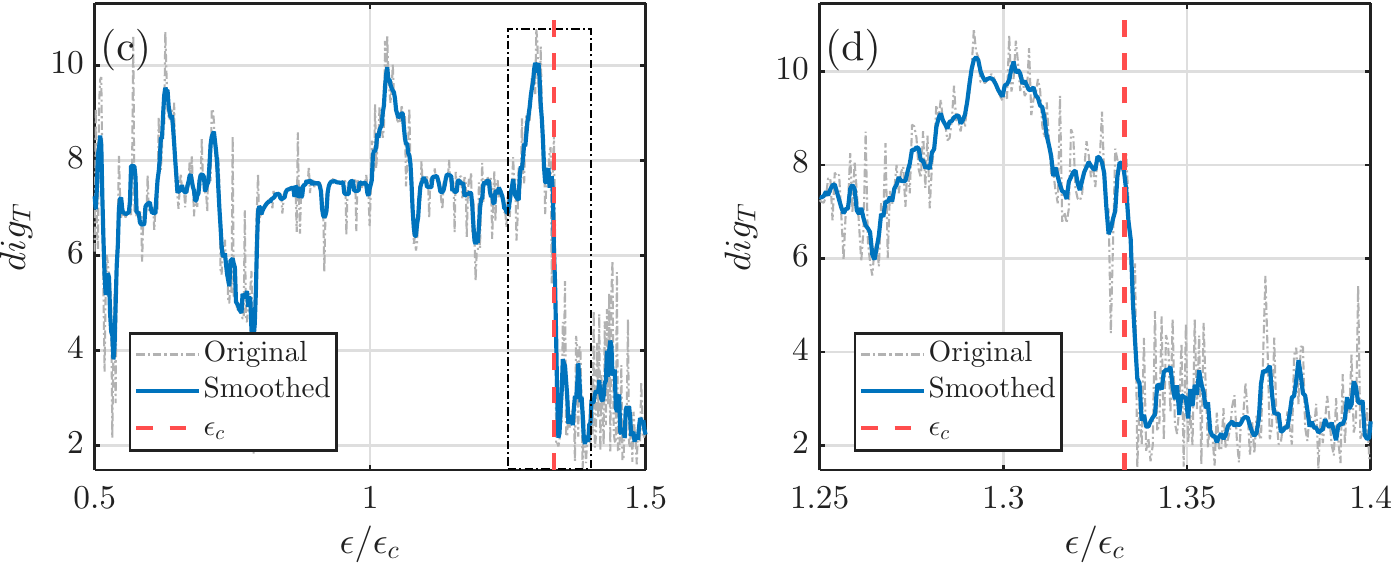}
\caption{
Comparison of the evolution of $\mathrm{dig}_T$ as a function of the normalized perturbation magnitude $\epsilon/\epsilon_c$ for two different MHD islands: (4,1) and (7,2).
(a, b) Results for the (4,1) MHD island: 
(a) global scan over a wide range of $\epsilon/\epsilon_c$ with the local region $0.85 \leq \epsilon/\epsilon_c \leq 0.95$ highlighted by a black dashed rectangle; 
(b) detailed view of the local region. 
(c, d) Results for the (7,2) MHD island: 
(c) global scan with the same conventions as (b) and (d) local scan with range $1.24 \leq \epsilon/\epsilon_c \leq 1.40$.
In all panels, both the original (dash-dot line) and smoothed (solid line) $\mathrm{dig}_T$ curves are shown. 
The vertical red dashed line indicates the drop point of $\mathrm{dig}_T$.
}
\label{fig:digT_vs_epsilon_all}
\end{figure}
% --------------- DIII-D_renormal_width_q ---------------
\clearpage
\begin{figure}[htbp]
\centering
\includegraphics[width=\linewidth]{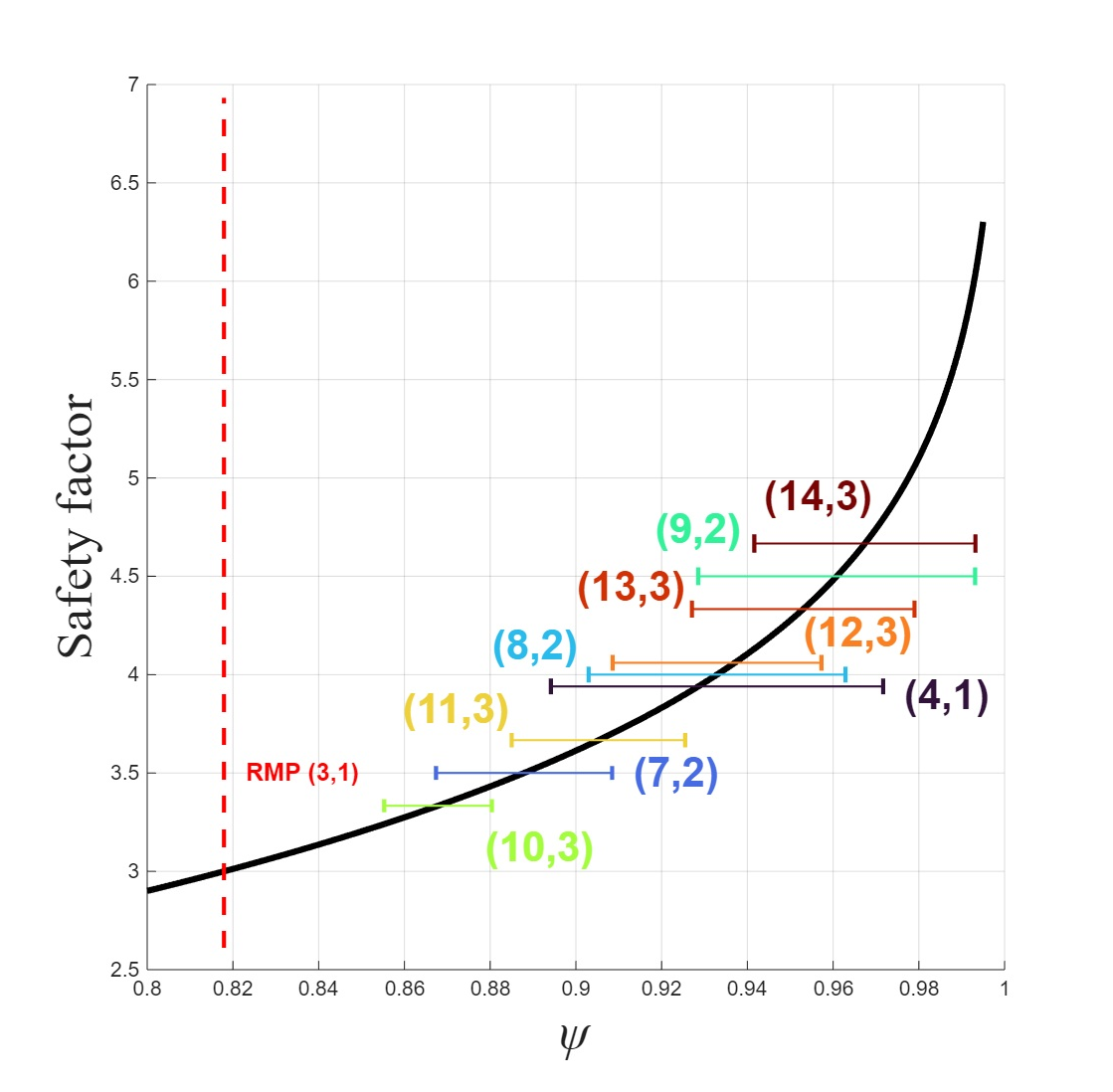}
\caption{The safety factor $q$-profile and the visualization of MHD magnetic island width. The black solid line represents the $q(\psi)$ profile. The horizontal bars of different colors indicate the critical magnetic islands excited by different MHD modes $(m,n)$ on their respective rational surfaces, with the horizontal length representing the magnetic island width $\Delta\psi$. When modes share the same rational surface (e.g., $q=4$), the horizontal bars are staggered in the vertical direction for distinction. The red dashed line indicates the position of the (3,1) RMP resonant magnetic surface.}
\label{fig:q_profile_islands}
\end{figure}
% --------------- DIII-D_Chirikov ---------------
\clearpage
\begin{figure}[htbp]
\centering
\includegraphics[width=\linewidth]{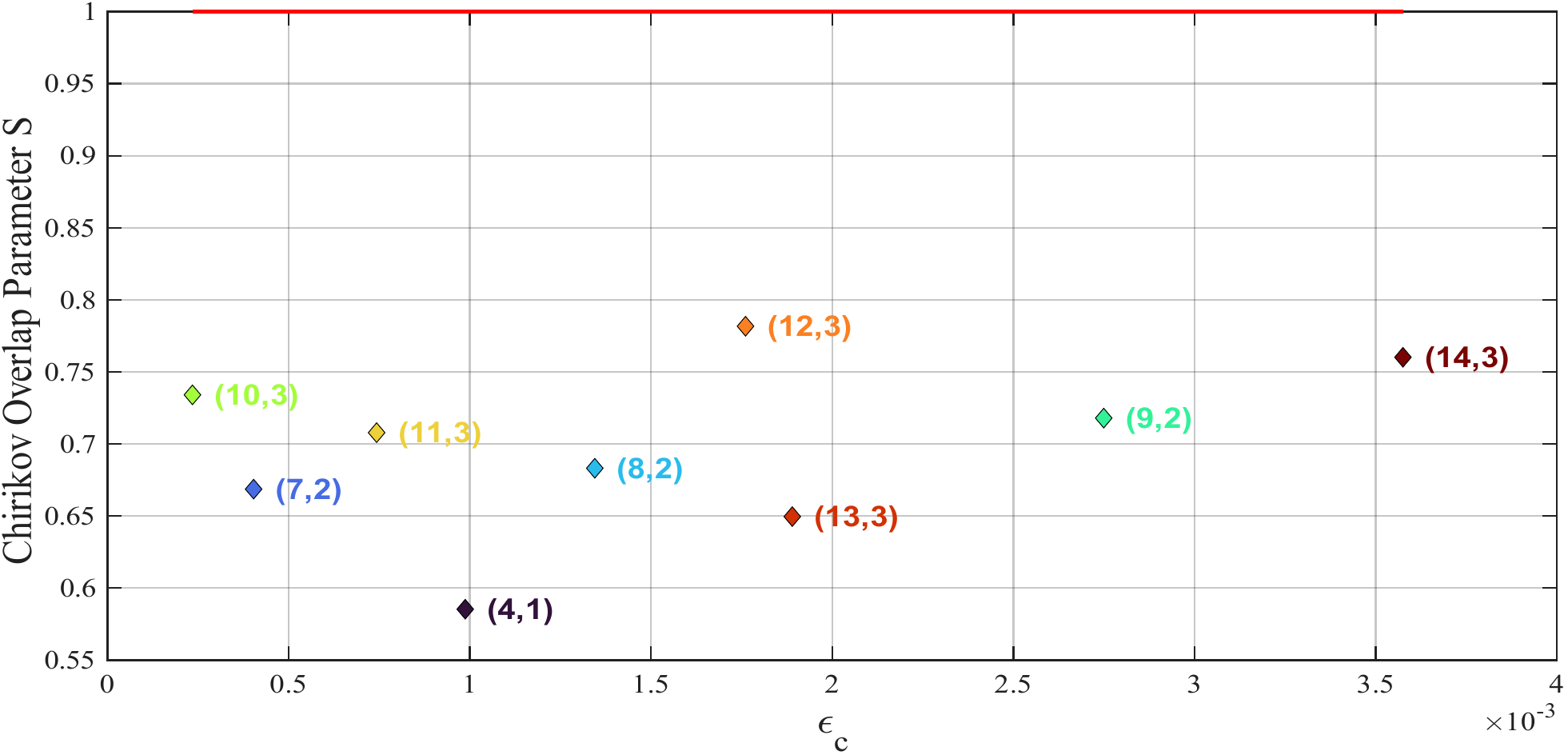}
\caption{Chirikov Overlap Parameter vs. Critical Epsilon. The red line indicates $S=1$}
\label{fig:Chirikov}
\end{figure}

% --------------- DIII-D_non_res ---------------
\clearpage
\begin{figure}[htbp]
\centering
    \begin{subfigure}[b]{\linewidth}
    \includegraphics[width=\linewidth]{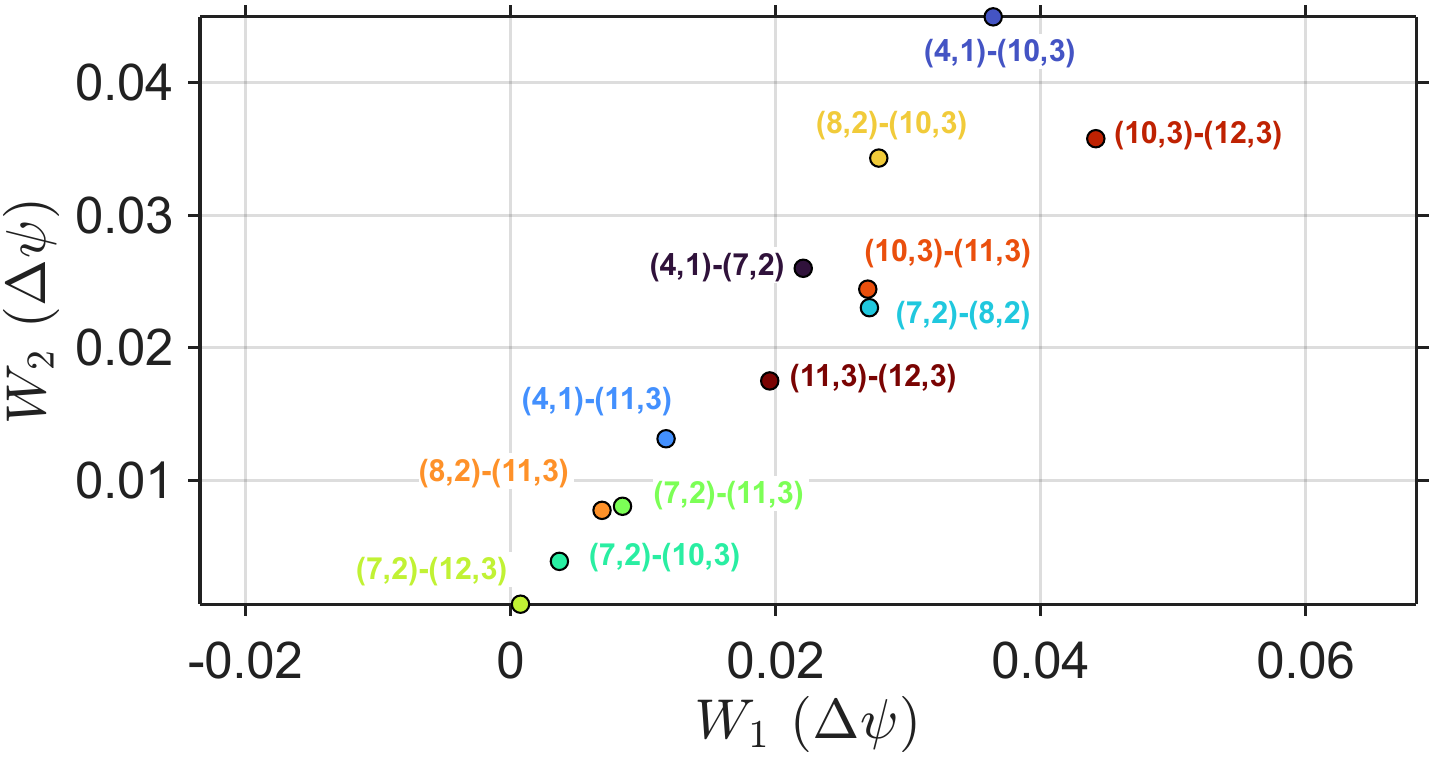}
    \end{subfigure}
    \hfill
    \begin{subfigure}[b]{\linewidth}
    \includegraphics[width=\linewidth]{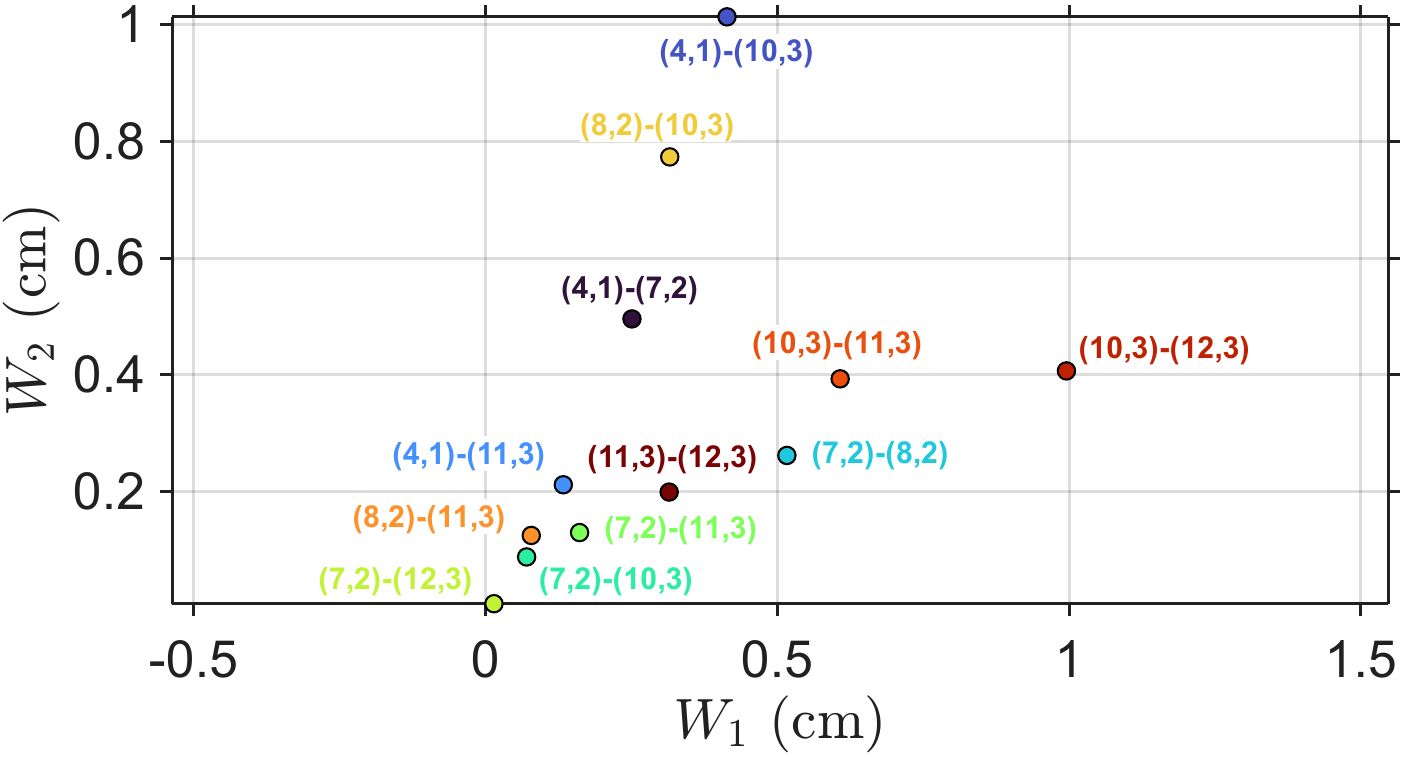}
    \end{subfigure}
\caption{Critical Magnetic Island Width: Comparison between pairs of non-resonant MHD islands. The upper subgraph shows the width relationship in the flux coordinate, while the down subgraph presents the same data in physical units $(\mathrm{cm})$. Each scatter represents a pair of edge MHD islands.}
\label{fig:nonresonant_MHD_scatter}
\end{figure}

\end{document}